\documentclass[preprint,12pt]{article}

  \usepackage{latexsym,bm,amsmath,amssymb,amsfonts,datetime}
  \usepackage{epsfig,graphics,graphicx}
  \usepackage{slashed}
  \usepackage{cite}
  \usepackage[latin1]{inputenc}
  \usepackage{mathrsfs}
\usepackage{mathcomp}
\usepackage{color}
\usepackage{caption2}

  \long\def\comment#1{ }

  \newcommand{\beq}{\begin{eqnarray}}
  \newcommand{\eeq}{\end{eqnarray}}
  
 \def\simge{\mathrel{%
   \rlap{\raise 0.511ex \hbox{$>$}}{\lower 0.511ex \hbox{$\sim$}}}}
\def\simle{\mathrel{
   \rlap{\raise 0.511ex \hbox{$<$}}{\lower 0.511ex \hbox{$\sim$}}}}

\newcommand{\nc}{\newcommand}

\usepackage[
      colorlinks=true,
      urlcolor=blue,    
      filecolor=blue,     
      citecolor=red,
      pdfstartview=FitV,
       bookmarksopen=true    
      ]{hyperref}

\definecolor{cardinal}{rgb}{0.6,0,0}
\definecolor{darkgreen}{rgb}{0,0.5,0}
\definecolor{golden}{rgb}{0.92, 0.7, 0}
\definecolor{midnight}{rgb}{0, 0, 0.5}
\definecolor{darkblue}{rgb}{0.2, 0, 0.8}

\nc{\ra}{\rightarrow} 
\nc{\lra}{\leftrightarrow} 
\nc{\Ra}{\Rightarrow} 
\nc{\LRa}{\Leftightarrow} 
\nc{\blp}{{\big (}}
\nc{\brp}{{\big )}}
\nc{\Blp}{{\Big (}}
\nc{\Brp}{{\Big )}}
\nc{\bglp}{{\bigg (}}
\nc{\bgrp}{{\bigg )}}
\nc{\Bglp}{{\Bigg (}}
\nc{\Bgrp}{{\Bigg )}}
\nc{\slb}{{\rm [}}
\nc{\srb}{{\rm ]}}
\nc{\bslb}{{\rm \big [}}
\nc{\bsrb}{{\rm \big ]}}
\nc{\Bslb}{{\rm \Big [}}
\nc{\Bsrb}{{\rm \Big ]}}
\def\al{\alpha}

\def\eps{\epsilon}
\nc{\veps}{\varepsilon}
\def\gam{\gamma}

\def\lam{\lambda}
\def\om{\omega}
\nc{\vphi}{\varphi}
\def\tha{\theta}
\def\sig{\sigma}

\def\Gam{\Gamma}

\def\Om{\Omega}
\def\Sig{\Sigma}

\nc{\myvspace}{\rule[-1em]{0pt}{2.5em}}
\nc{\bea}{\begin{eqnarray}}
\nc{\eea}{\end{eqnarray}}
\nc{\be}{\begin{equation}}
\nc{\ee}{\end{equation}}
\nc{\barr}{\begin{array}}
\nc{\earr}{\end{array}}
\nc{\cA}{{\cal A}}
\nc{\cB}{ \cal B}

\def\cD{{\cal D}}

\nc{\cF}{{\cal F}}
\nc{\cG}{{\cal G}}

\nc{\cL}{{\cal L}}
\nc{\cM}{{\cal M}}

\def\cO{{\cal O}}

\nc{\cQ}{{\cal Q}}
\nc{\cR}{{\cal R}}

\def\cV{{\cal V}}
\def\cV{{\cal V}}

\def\cZ{{\cal Z}}
\nc{\cQd}{\cQ^{\dagger}}
\nc{\cRd}{\cR^{\dagger}}
\nc{\BB}{{\mathbb B}}
\nc{\CC}{{\mathbb C}}
\nc{\DD}{{\mathbb D}}
\nc{\EE}{{\mathbb E}}
\nc{\FF}{{\mathbb F}}
\nc{\GG}{{\mathbb G}}
\nc{\HH}{{\mathbb H}}
\nc{\JJ}{{\mathbb J}}
\nc{\RR}{{\mathbb R}}
\nc{\PP}{{\mathbb P}}
\nc{\QQ}{{\mathbb Q}}
\nc{\ZZ}{{\mathbb Z}}
\nc{\calone}{{\mathbb 1}}

\nc{\half}{\frac{1}{2}}
\nc{\qrt}{\frac{1}{4}}

\nc{\delbar}{\bar\partial}
\nc{\thalf}{\frac{t}{2}}
\nc{\Spin}{\operatorname{Spin}}
\nc{\SO}{\operatorname{SO}}

\nc{\Sp}{{\rm Sp}}
\nc{\com}[2]{{ \left[ #1, #2 \right] }}
\nc{\acom}[2]{{ \left\{ #1, #2 \right\} }}
\nc{\rr}{\rightarrow}
\nc{\p}{\partial}
\nc{\LT}{{\LL_\T}}
\def\com#1#2{{ \left[ #1, #2 \right] }}
\def\acom#1#2{{ \left\{ #1, #2 \right\} }}
\nc{\Adag}{A^{\dagger}}
\nc{\AdagI}{A^{\dagger I}}
\nc{\AdagJ}{A^{\dagger J}}
\nc{\AdagK}{A^{\dagger K}}
\nc{\AdagL}{A^{\dagger L}}
\nc{\AdagM}{A^{\dagger M}}
\nc{\Bdag}{B^{\dagger}}
\nc{\BdagI}{B^{\dagger}_I}
\nc{\BdagJ}{B^{\dagger}_J}
\nc{\BdagK}{B^{\dagger}_K}
\nc{\BdagL}{B^{\dagger}_L}
\nc{\BdagM}{B^{\dagger}_M}
\nc{\Cdag}{C^{\dagger}}
\nc{\CdagI}{C^{\dagger I}}
\nc{\CdagJ}{C^{\dagger J}}
\nc{\CdagK}{C^{\dagger K}}
\nc{\Ddag}{D^{\dagger}}
\nc{\DdagI}{D^{\dagger I}}
\nc{\DdagJ}{D^{\dagger J}}
\nc{\DdagK}{D^{\dagger K}}
\nc{\ttha}{\tilde{\theta}}
\nc{\tphi}{\tilde{\phi}}
\nc{\tsig}{\widetilde{\sig}}
\nc{\tom}{\tilde{\om}}
\nc{\tlam}{\tilde{\lam}}
\nc{\tSig}{\widetilde{\Sig}}
\nc{\tPhi}{\tilde{\Phi}}
\nc{\tPhibar}{\ol{\tPhi}}
\nc{\tPi}{\tilde{\Pi}}
\nc{\tpsi}{\tilde{\psi}}
\nc{\tPsi}{\tilde{\Psi}}
\nc{\tgam}{\tilde{\gam}}
\nc{\tGam}{\tilde{\Gam}}
\nc{\tzeta}{\tilde{\zeta}}
\nc{\tZeta}{\tilde{\Zeta}}
\nc{\teta}{\tilde{\eta}}
\nc{\teps}{\tilde{\eps}}
\nc{\tEta}{\tilde{\Eta}}
\nc{\tchi}{\tilde{\chi}}
\nc{\tChi}{\tilde{\Chi}}
\nc{\txi}{\tilde{\xi}}
\nc{\tXi}{\tilde{\Xi}}
\nc{\tb}{\tilde b}
\nc{\tc}{\tilde c}
\nc{\te}{\tilde e}
\nc{\tf}{\tilde f}
\nc{\tg}{\tilde g}
\nc{\tj}{\tilde j}
\nc{\tp}{\widetilde{p}}
\nc{\tq}{\widetilde{q}}
\nc{\ts}{{\tilde s}}
\nc{\tu}{{\tilde u}}
\nc{\tv}{{\tilde v}}
\nc{\tw}{{\tilde w}}
\nc{\tx}{{\tilde x}}
\nc{\ty}{{\tilde y}}
\nc{\tz}{\tilde z}
\nc{\tA}{{\widetilde A}}
\nc{\tAbar}{{\ol \tA}}
\nc{\tB}{{\widetilde B}}
\nc{\tC}{{\widetilde C}}
\nc{\tD}{{\widetilde D}}
\nc{\tE}{{\widetilde E}}
\nc{\tG}{{\widetilde G}}
\nc{\tH}{{\widetilde H}}
\nc{\tJ}{{\widetilde J}}
\nc{\tJbar}{{\ol {\tilde J}}}
\nc{\tK}{{\widetilde K}}
\nc{\tL}{{\widetilde L}}
\nc{\tM}{{\widetilde M}}
\nc{\tN}{{\widetilde N}}
\nc{\tP}{{\widetilde P}}
\nc{\tQ}{{\widetilde Q}}
\nc{\tR}{{\widetilde R}}
\nc{\tS}{\widetilde{S}}
\nc{\tF}{\tilde{{\cal F}}}
\nc{\tX}{\widetilde{X}}
\nc{\tY}{\widetilde{Y}}
\nc{\tcZ}{\tilde{\cZ}}
\nc{\tcZbar}{\ol{\tcZ}}
\nc{\hb}{\hat b}
\nc{\hc}{\hat c}
\nc{\hd}{\hat d}
\nc{\he}{\hat e}
\nc{\hf}{\hat f}
\nc{\hg}{\hat g}
\nc{\hh}{\hat h}
\nc{\hp}{\hat p}
\nc{\hs}{\hat s}
\nc{\hv}{\hat v}
\nc{\hw}{\hat w}
\nc{\hx}{\hat x}
\nc{\hy}{\hat y}
\nc{\hz}{\hat z}
\nc{\zhat}{\hat z}
\nc{\hA}{\widehat{A}}
\nc{\hE}{\widehat{E}}
\nc{\hF}{\widehat{F}}
\nc{\hH}{\widehat{H}}
\nc{\hJ}{\widehat{J}}
\nc{\hK}{\widehat{K}}
\nc{\hL}{\widehat{L}}
\nc{\hM}{\widehat M}
\nc{\hN}{\widehat{N}}
\nc{\hV}{\widehat V}
\nc{\hcV}{\widehat \cV}

\nc{\ha}{\widehat \alpha}
\nc{\hphi}{\hat{\phi}}
\nc{\hpsi}{\hat{\psi}}
\nc{\hgam}{\hat{\gam}}
\nc{\hPhi}{\hat{\Phi}}
\nc{\hPsi}{\hat{\Psi}}
\nc{\hGam}{\hat{\Gam}}
\nc{\omhat}{\hat{\om}}
\nc{\hOm}{\widehat{\Om}}

\nc{\w}{\wedge}

\nc{\vb}{\vec b}
\nc{\vc}{\vec c}
\nc{\vd}{\vec d}
\nc{\ve}{\vec e}
\nc{\vf}{\vec f}
\nc{\vg}{\vec g}
\nc{\vh}{\vec h}
\nc{\vp}{\vec p}
\nc{\vq}{\vec q}
\nc{\vr}{\vec r}
\nc{\vs}{\vec s}
\nc{\vv}{\vec v}
\nc{\vw}{\vec w}
\nc{\vx}{\vec x}
\nc{\vy}{\vec y}
\nc{\vz}{\vec z}

\nc{\vB}{\vec B}
\nc{\vC}{\vec C}
\nc{\vD}{\vec D}
\nc{\vE}{\vec E}
\nc{\vF}{\vec F}
\nc{\vG}{\vec G}
\nc{\vH}{\vec H}
\nc{\vP}{\vec P}
\nc{\vQ}{\vec Q}
\nc{\vR}{\vec R}
\nc{\vS}{\vec S}
\nc{\vV}{\vec V}
\nc{\vW}{\vec W}
\nc{\vX}{\vec X}
\nc{\vY}{\vec Y}
\nc{\vZ}{\vec Z}

\nc{\ol}{\overline}
\nc{\bbar}{\ol{b}}
\nc{\cbar}{\ol{c}}
\nc{\dbar}{\ol{d}}
\nc{\ebar}{\ol{e}}
\nc{\ibar}{\ol{\imath}}
\nc{\jbar}{\ol{\jmath}}
\nc{\kbar}{\ol{k}}
\nc{\lbar}{\ol{l}}
\nc{\mbar}{\ol{m}}
\nc{\nbar}{\ol{n}}
\nc{\pbar}{\ol{p}}
\nc{\qbar}{\ol{q}}
\nc{\ubar}{\ol{u}}
\nc{\vbar}{\ol{v}}
\nc{\wbar}{\ol{w}}
\nc{\xbar}{\ol{x}}
\nc{\ybar}{\ol{y}}
\nc{\zbar}{\ol{z}}

\nc{\Abar}{\ol{A}}
\nc{\Bbar}{\ol{B}}
\nc{\Cbar}{\ol{C}}
\nc{\Dbar}{\ol{D}}
\nc{\Ebar}{\ol{E}}
\nc{\Fbar}{\ol{F}}
\nc{\Hbar}{\ol{H}}
\nc{\Jbar}{\ol{J}}
\nc{\Kbar}{\ol{K}}
\nc{\Lbar}{\ol{L}}
\nc{\Mbar}{\ol{M}}
\nc{\Nbar}{\ol{N}}
\nc{\Pbar}{\ol{P}}
\nc{\Qbar}{\ol{Q}}
\nc{\Rbar}{\ol{R}}
\nc{\Sbar}{\ol{S}}
\nc{\Tbar}{\ol{T}}
\nc{\Ubar}{\ol{U}}
\nc{\Vbar}{\ol{V}}
\nc{\Wbar}{\ol{W}}
\nc{\Xbar}{{\overline X}}
\nc{\Ybar}{{\overline Y}}
\nc{\Zbar}{{\overline Z}}
\nc{\cZbar}{{\overline \cZ}}

\nc{\epsbar}{\ol{\epsilon}}
\nc{\lambar}{\ol{\lambda}}
\nc{\zetabar}{\ol{\zeta}}
\nc{\Zetabar}{\ol{\Zeta}}
\nc{\psibar}{\ol{\psi}}
\nc{\Psibar}{\ol{\Psi}}
\nc{\phibar}{\ol{\phi}}
\nc{\Phibar}{\ol{\Phi}}
\nc{\chibar}{\ol{\chi}}
\nc{\mubar}{\ol{\mu}}
\nc{\nubar}{\ol{\nu}}
\nc{\rhobar}{\ol{\rho}}
\nc{\ombar}{\ol{\om}}
\nc{\Ombar}{\ol{\Om}}
\nc{\Deltabar}{\ol{\Delta}}
\nc{\Thetabar}{\ol{\Theta}}
\nc{\xibar}{\ol{\xi}}
\nc{\Xibar}{\ol{\Xi}}

\nc{\Dthbar}{\ol{\rm D3}}

\nc{\gdot}{\dot{g}}
\nc{\xdot}{\dot{x}}
\nc{\ydot}{\dot{y}}
\nc{\phidot}{\dot{\phi}}
\nc{\sinp}{s_{\phi}}
\nc{\cosp}{c_{\phi}}
\nc{\tanp}{t_{\phi}}
\nc{\spone}{s_{\phi_1}}
\nc{\cpone}{c_{\phi_1}}
\nc{\tpone}{t_{\phi_1}}
\nc{\sptwo}{s_{\phi_2}}
\nc{\cptwo}{c_{\phi_2}}
\nc{\tptwo}{t_{\phi_2}}
\nc{\spth}{s_{\phi_3}}
\nc{\cpth}{c_{\phi_3}}
\nc{\tpth}{t_{\phi_3}}
\nc{\calp}{c_{\al}}
\nc{\salp}{s_{\al}}

\nc{\csch}{{\rm csch}}
\nc{\sech}{{\rm sech}}

\nc{\cothzlami}{\coth(z-\lam_i)}
\nc{\coshzlami}{\cosh(z-\lam_i)}
\nc{\sinhzlami}{\sinh(z-\lam_i)}

\nc{\cothzlamj}{\coth(z-\lam_j)}
\nc{\coshzlamj}{\cosh(z-\lam_j)}
\nc{\sinhzlamj}{\sinh(z-\lam_j)}

\nc{\cothlamij}{\coth(\lam_i-\lam_j)}
\nc{\coshlamij}{\cosh(\lam_i-\lam_j)}
\nc{\sinhlamij}{\sinh(\lam_i-\lam_j)}
\nc{\bah}{{\mathbf {\hat{A}}}}
\nc{\bX}{{\mathbf X}}
\nc{\ba}{{\bf a}}
\nc{\bb}{{\bf b}}
\nc{\bc}{{\bf c}}
\nc{\bd}{{\bf d}}
\nc{\bg}{{\bf g}}
\nc{\bk}{{\bf k}}
\nc{\bl}{{\bf l}}
\nc{\bn}{{\bf n}}
\nc{\bo}{{\bf o}}
\nc{\bp}{{\bf p}}
\nc{\bq}{{\bf q}}
\nc{\br}{{\bf r}}
\nc{\bs}{{\bf s}}
\nc{\bt}{{\bf t}}
\nc{\bu}{{\bf u}}
\nc{\bv}{{\bf v}}
\nc{\bw}{{\bf w}}
\nc{\bx}{{\bf x}}
\nc{\by}{{\bf y}}
\nc{\bz}{{\bf z}}
\nc{\bom}{{\bf \om}}
\nc{\bombar}{{\mathbf \ombar}}
\nc{\bPhi}{{\bf \Phi}}

\nc{\rma}{{\rm a}}
\nc{\rmb}{{\rm b}}
\nc{\rmc}{{\rm c}}
\nc{\rmg}{{\rm g}}
\nc{\rk}{{\rm k}}
\nc{\rml}{{\rm l}}
\nc{\rmm}{{\rm m}}
\nc{\rmn}{{\rm n}}
\nc{\rmo}{{\rm o}}
\nc{\rmp}{{\rm p}}
\nc{\rmq}{{\rm q}}
\nc{\rmr}{{\rm r}}
\nc{\rms}{{\rm s}}
\nc{\rmt}{{\rm t}}
\nc{\rmu}{{\rm u}}
\nc{\rmv}{{\rm v}}
\nc{\rmw}{{\rm w}}
\nc{\rmx}{{\rm x}}
\nc{\rmy}{{\rm y}}
\nc{\rmz}{{\rm z}}
\nc{\Ffour}{{F^{(4)}}}
\nc{\Ffive}{{F^{(5)}}}

\nc{\dal}{\dot{\al}}
\nc{\thadot}{\dot{\tha}}
\nc{\thab}{\bar{\theta}}
\nc{\thal}{\theta^{\al}}
\nc{\thdal}{\bar{\theta}^{\dal}}

\nc{\thsigthm}{\tha \sigma^m \thab}
\nc{\thsigthn}{\tha \sigma^n \thab}

\nc{\Dal}{D_{\al}}
\nc{\Ddal}{\bar{D}_{\dal}}
\nc{\CDal}{{\cal D}_{\al}}
\nc{\CDdal}{\bar{\cal D}_{\dal}}

\nc{\eq}[1]{(\ref{#1})}
\nc{\non}{\nonumber}

\nc{\equ}{{\rm eq}}

\nc{\vol}{{\rm vol}}
\nc{\Ainf}{A_{\infty}}
\nc{\End}{{\rm End}}
\nc{\Ext}{{\rm Ext}}
\nc{\IIB}{{\rm IIB}}
\nc{\Ad}{{\rm Ad}}
\nc{\IIA}{{\rm IIA}}
\nc{\AdS}{{\rm AdS}}
\nc{\CFT}{{\rm CFT}}
\nc{\Dslash}{\ensuremath \raisebox{0.025cm}{\slash}\hspace{-0.32cm} D}
\nc{\cDslash}{\ensuremath \raisebox{0.025cm}{\slash}\hspace{-0.32cm} \cD}
\nc{\no}{\!:\!\!}
\nc{\ointdz}{\oint\frac{dz}{2\pi i}}
\nc{\ointdzone}{\oint\frac{dz_1}{2\pi i}}
\nc{\ointdztwo}{\oint\frac{dz_2}{2\pi i}}
\nc{\ointdzb}{\oint\frac{d\zbar}{2\pi i}}
\nc{\ointdzbone}{\oint\frac{d\zbar_1}{2\pi i}}
\nc{\ointdzbtwo}{\oint\frac{d\zbar_2}{2\pi i}}
\nc{\dz}{\frac{dz}{2\pi i}}
\nc{\dzb}{\frac{d\zbar}{2\pi i}}
\nc{\bpm}{\begin{pmatrix}}
\nc{\epm}{\end{pmatrix}}
 \nc{\bitem}{\begin{itemize}}
 \nc{\eitem}{\end{itemize}}

\begin{document}
\begin{flushright}
IPhT-T10/169
\end{flushright}
\vspace{0.5cm}
\begin{center}
 \vskip1.5cm 

\baselineskip=13pt {\LARGE \bf{The Backreaction of Anti--M2 Branes \\ on a Warped Stenzel Space}\\}
 \vskip1.5cm 
 Iosif Bena$^{\dagger}$, Gregory Giecold$^{\dagger}$ and Nick Halmagyi$^{\dagger *}$\\ 
 \vskip0.5cm
$^{\dagger}$\textit{Institut de Physique Th\'eorique,\\
CEA Saclay, CNRS URA 2306,\\
F--91191 Gif--sur--Yvette, France}\\
\vskip0.8cm
$^{*}$\textit{Laboratoire de Physique Th\'eorique et Hautes Energies,\\
Universit\'e Pierre et Marie Curie, CNRS UMR 7589, \\
F--75252 Paris Cedex 05, France}\\
\vskip0.5cm
iosif.bena@cea.fr, gregory.giecold@cea.fr \\
halmagyi@lpthe.jussieu.fr \\ 

\end{center}
\vspace{1.0cm}

\abstract{We find the superpotential governing the supersymmetric
  warped M--theory solution with a transverse Stenzel space found by
  Cveti\v c, Gibbons, L\"u and Pope in hep--th/0012011, and use this
  superpotential to extract and solve the twelve coupled equations
  underlying the first--order backreacted solution of a stack of
  anti--M2 branes in this space. These anti--M2 branes were analyzed
  recently in a probe approximation by Klebanov and Pufu, who
  conjectured that they should be dual to a metastable vacuum of a
  supersymmetric 2+1 dimensional theory.  We find that the would--be
  supergravity dual to such a metastable vacuum must have an infrared
  singularity and discuss whether this singularity is acceptable or
  not. Given that a similar singularity appears when placing anti--D3
  branes in the Klebanov--Strassler solution, our work strengthens the
  possibility that anti--branes in warped throats do not give rise to
  metastable vacua.}

\newpage

\section{Introduction and discussion}
\setcounter{equation}{0}

The recent revival of interest in metastable supersymmetry breaking in
quantum field theory is largely due to the work of Intriligator,
Seiberg and Shih~\cite{Intriligator:2006dd} (ISS).  This work presents
a mechanism to naturally circumvent some of the problems afflicting
other models for dynamic supersymmetry breaking
(DSB)~\cite{Shadmi:1999jy, Nelson:1993nf, Intriligator:2007py,
  Intriligator:2007cp}. A natural question that was posed immediately
after~\cite{Intriligator:2006dd} is whether metastable vacua also
exist in string realizations of supersymmetric field theories.

For type IIA brane--engineering models of supersymmetric field
theories, the answer to this question is negative~\cite{Bena:2006rg}.
Indeed, these models are constructed using D4 branes ending on
codimension--two defects inside NS5
branes~\cite{Ooguri:2006bg,Franco:2006ht,Bena:2006rg}, which source
NS5 worldvolume fields that grow logarithmically at infinity. In
supersymmetric vacua this logarithmic growth encodes the running of
the gauge theory coupling constant with the
energy~\cite{Witten:1997sc, Hori:1997ab, Brandhuber:1997iy,
  Giveon:1998sr}, but these logarithmic modes are different in the
candidate metastable brane configuration and in the supersymmetric
one. This implies that the candidate metastable brane configuration
and the supersymmetric one differ by an infinite amount, and hence
cannot decay into each other.  Hence, the type IIA brane construction
does not describe a metastable vacuum of a supersymmetric theory, but
instead a nonsupersymmetric vacuum of a nonsupersymmetric theory.

Another arena where one might try to find string theory realizations of
metastable vacua are IIB holographic duals of certain supersymmetric gauge theories.
The best--known example in this class was proposed by Kachru, Pearson
and Verlinde~\cite{Kachru:2002gs, DeWolfe:2004qx}, who argued that a background with
anti--D3 branes at the bottom of the Klebanov--Strassler warped deformed
conifold~\cite{Klebanov:2000hb} is dual to a metastable vacuum of the
dual supersymmetric gauge theory. Since the Klebanov--Strassler
solution has positive D3 brane charge dissolved in flux, the anti--D3
branes can annihilate against this charge (this annihilation happens
via the polarization of the anti--D3 branes into an NS5 brane~\cite{Myers:1999ps,Polchinski:2000uf}), and this bulk process
is argued to correspond to the decay of the metastable vacuum to the
supersymmetric one in the dual field theory.

Another proposal for a metastable vacuum obtained by putting
anti--branes at the bottom of a smooth warped throat with positive
brane charge dissolved in flux has recently been made by Klebanov and
Pufu~\cite{Klebanov:2010qs}, who argued that probe anti--M2 branes at
the tip of a supersymmetric warped M--theory background with
transverse Stenzel space~\cite{Stenzel:1993}, give rise to a
long--lived metastable vacuum. The supersymmetric solution, first
found by Cveti\v c, Gibbons, L\"u and Pope (CGLP)
in~\cite{Cvetic:2000db} has M2 charge dissolved in fluxes and a large
$S^4$ in the infrared. The anti--branes can annihilate against the
charge dissolved in fluxes by polarizing into M5 branes~\cite{Bena:2000zb} 
wrapping three--spheres inside the $S^4$.

The probe brane analyses described above, while indicative that a
metastable vacuum might exist, are however not enough to establish
this.  One possible issue which can cause the backreacted solution to
differ significantly from the probe analysis is the presence of
non-normalizable modes. If the anti-branes indeed source such modes
then the candidate metastable configuration is not dual to a
non--supersymmetric vacuum of a supersymmetric theory, but to a
non--supersymmetric vacuum of a non--supersymmetric theory, and the
supersymmetry breaking is not dynamical but explicit.  The existence
of non--normalizable modes is not visible in the probe approximation
(much like the existence of type IIA log--growing modes was not
visible in $g_s=0$ brane
constructions~\cite{Ooguri:2006bg,Franco:2006ht}), but only upon
calculating the backreaction of the probe branes -- a not too easy
task.

In~\cite{Bena:2009xk} two of the authors and M.~Gra\~ na found the
possible first--order backreacted solution sourced by a stack of
anti--D3 branes smeared on the large $S^3$ at the bottom of the
Klebanov--Strassler (KS) solution, and found two very interesting
features: first, of the 14 physical modes describing $SU(2) \times
SU(2) \times Z_2$--invariant perturbations of the warped deformed
conifold, only one mode enters in the expression of the force that a
probe D3 brane feels in this background. Hence, since anti--D3 branes
attract probe branes, if the perturbed solution is to have any chance
to describe backreacted anti--D3 branes, this mode must be
present\footnote{The asymptotic behavior of the force matches the one
  argued for in~\cite{Kachru:2003sx}, and the existence of this mode
  was first intuited in~\cite{DeWolfe:2008zy} which set out to study
  the UV asymptotics of the perturbations corresponding to anti--D3
  branes in the KT background\cite{Klebanov:2000nc}.}. The second
feature of this solution is that if the force mode is present, the
infrared\footnote{An IR analysis of some of the non--supersymmetric
  isometry--preserving perturbations of the Klebanov--Strassler
  background can also be found in~\cite{McGuirk:2009xx}.} must contain a certain singularity, 
  which has {\it finite} action\footnote{This was first observed by I. Klebanov.}. Note that
having a finite action does not automatically make a singularity
acceptable -- negative--mass Schwarzschild is an obvious
counterexample~\cite{Horowitz:1995ta}. As discussed
in~\cite{Bena:2009xk}, if this singularity is unphysical, then the
solution sourced by the anti--D3 branes cannot be thought of as a
small perturbation of the KS solution, and therefore does not describe
a metastable vacuum of the dual theory. If this singularity is
physical, the first--order solution does describe anti--D3 branes at
the bottom of the KS solution, and work is in progress to determine
what are the features of this solution, and whether the perturbative
anti--D3 brane solution describes or not metastable vacua of the dual
theory.

The purpose of this paper is to calculate the first--order
backreaction of the other proposed metastable configuration with
anti--branes in a background with charge dissolved in fluxes: the
anti--M2 branes in the Stenzel--CGLP solution~\cite{Cvetic:2000db}. In
order to do this we smear the anti--M2 branes on the large $S^4$ at
the bottom of the Stenzel--CGLP solution, and solve for all possible
deformations of this background that preserve its $\text{SO}(5)$
symmetry. We consider an ansatz for these deformations ; the space of
deformations is parameterized by 6 functions of one variable
satisfying second--order differential equations. However, when
perturbing around a supersymmetric solution, Borokhov and
Gubser~\cite{Borokhov:2002fm} have observed that these second--order
equations factorize into first--order ones, that are much easier to
solve. Nevertheless, in order to apply the Borokhov--Gubser method,
one needs to find the superpotential underlying the supersymmetric
solution, which for the warped fluxed Stenzel--CGLP solution was not
known until now. The first result of this paper, presented in Section 2, is to
find this superpotential\footnote{This is the equivalent of the
  Papadopoulos--Tseytlin superpotential for the KS
  solution~\cite{Papadopoulos:2000gj, Cassani:2010na, Bena:2010pr}.},
and derive two sets of first--order equations governing the space of
deformations.
 
We then show in Section 3 that the force felt by a probe M2 brane in
the most general perturbed background depends on only one of the
``conjugate--momentum'' functions that appear when solving the
first--order system, and hence on only one of the 10 constants
parameterizing the deformations around the supersymmetric solution. We
then solve in Section 4 the two sets of first--order differential
equations.  Amazingly enough, the solutions for the first set of
equations (for the conjugate--momentum functions) can be found
explicitly in terms of incomplete elliptic integrals (a huge
improvement on the situation in~\cite{Bena:2009xk}). We also find the
homogeneous solutions to the other equations and give implicitly the
full solution to the system in terms of integrals. We also provide the
explicit UV and IR expansions of the full space of deformations, and
find which deformations correspond to normalizable modes and which
deformations correspond to non--normalizable modes.

In Section 5 we then use the machinery we developed to recover the
perturbative expansion of the known solution sourced by BPS M2 branes
smeared on the $S^4$ at the tip of the Stenzel--CGLP
solution~\cite{Cvetic:2000db}, and analyze the infrared of the
possible solution sourced by anti--M2 branes. After removing some
obviously unphysical divergences and demanding that in the
first--order backreacted solution a probe M2 brane feels a nonzero
force, we find that the only backreacted solution that can correspond
to anti--M2 branes must have an infrared singularity, coming from a
four--form field strength with two or three legs on the three--sphere
that is shrinking to zero size at the tip of the Stenzel space.

Hence, the first--order backreacted solution for the anti--M2 branes
has the same two key features as the anti--D3 branes in KS: the force
felt by a probe M2 brane in this background depends only on one of the
10 physical perturbation modes around this solution, and the solution
where the force--carrying mode is turned on must have an infrared
singularity coming from a divergent energy in the M--theory four--form
field strength. Nevertheless, unlike in the ``anti--D3 in KS"
solution, the action of this infrared singularity also diverges.
Again, if this singularity is physical, our first--order backreacted
solution describes anti--M2 branes in the CGLP background, and, to our
knowledge, would be the first backreacted supergravity solution dual
to metastable susy--breaking in 2+1 dimensions since the work of Maldacena and N\u
astase~\cite{Maldacena:2001pb}. This may be of interest both in the
programme of using the AdS/CFT correspondence to describe
strongly-interacting condensed--matter systems, and
also in view of the relevance of three--dimensional QFT's at strong
coupling to a recent holographic model of four--dimensional
cosmology~\cite{McFadden:2010na}. On the other hand, if the
singularity is not physical then the backreaction of the anti--M2
branes cannot be taken into account perturbatively; this indicates
that the only solution with proper anti--M2 brane boundary conditions
in the infrared is the solution for anti--M2 branes in a CGLP
background with anti--M2 brane charge dissolved in flux, and hence the
anti--M2 branes flip the sign of the M2 brane charge dissolved in flux.

Given the similarity of the results of the ``anti--D3 in KS'' and of
the ``anti--M2 in CGLP'' analyses and the drastically--different
calculations leading to them, it is rather natural to expect that the
underlying physics of the two setups is the same: either both
singularities are physical, which indicates that anti--branes in
backgrounds with charge dissolved in fluxes give rise to metastable
vacua, or they are both unphysical, which supports the idea that
anti--branes in such backgrounds cannot be treated as a perturbation of the original
solution, and may flip the sign of the charge dissolved
in flux. Furthermore, our analysis suggests that one cannot use the
finiteness of the action as a criterion for accepting a singularity.
This would allow the anti--D3 singularity and exclude the anti--M2
one, which would be rather peculiar, given the striking resemblance of
the two systems.

There are a few possible explanations for the singularities we
encounter in the anti--M2 and anti--D3 solutions.  One is that these singularities are accompanied by
stronger, physical singularities, coming from the smeared anti--M2 or anti--D3
sources, and one can hope that whatever mechanism renders the stronger
singularities physical may cure the subleading ones as well. Another
explanation is that the subleading singularities are a result of
smearing the antibranes. This is a difficult argument to support with
calculational evidence, as the unsmeared solution is a formidable
problem even for BPS branes in Stenzel spaces \cite{Krishnan:2008gx,
  Pufu:2010ie}.  Furthermore, a naive comparison of the anti--M2 and
anti--D3 solutions indicates that the stronger the physical singularity
associated with the brane sources is, the stronger the subleading
singularity will be. Hence, it is likely that unsmearing will make
things worse, not better. Note also that one cannot link the divergent
four--form field strength with the M5 branes into which the anti--M2
branes at the tip of the Stenzel--CGLP solution polarize -- they have
incompatible orientations.

It is also interesting to remember that when one attempts to build
string realisations of four-dimensional metastable vacua, either via
brane constructions~\cite{Bena:2006rg} or via
AdS-CFT~\cite{Bena:2009xk}, the non--normalizable modes one encounters are
log--growing modes, which one could in hindsight have expected 
from the generic running of coupling constants of four--dimensional 
gauge theories with the energy. 

For anti--M2 branes there is no such link. There exist both AdS/CFT
duals of metastable vacua of 2+1 dimensional gauge
theories~\cite{Maldacena:2001pb}, as well as brane--engineering
constructions of such metastable vacua (using D3 branes ending on
codimension--three defects inside NS5 branes)~\cite{Giveon:2009bv}.
The nonexistence of an anti--M2 metastable vacuum could only be seen
in supergravity, and comes from the way the fields of the anti--M2
brane interact with the magnetic fields that give rise to the charge
dissolved in fluxes. This may indicate there is a problem with trying
to construct metastable vacua in string theory by putting antibranes
in backgrounds with charge dissolved in fluxes.  In an upcoming
paper~\cite{Giecold:D2} we will also argue that anti--D2 branes in
backgrounds with D2 brane charge dissolved in
fluxes~\cite{Cvetic:2001ma}, that one of us investigated
in~\cite{Giecold:2009wj}, have similar problems.

\section{Perturbations around a supersymmetric solution}
\setcounter{equation}{0}

We are interested in the backreaction of a set of anti--M2 branes
spread on a four--sphere at the bottom of the warped Stenzel
geometry~\cite{Stenzel:1993} with nontrivial fluxes. Smearing the
anti--M2's is necessary in order for the perturbed solution to have
the same $\text{SO}(5)$ global symmetry as the supersymmetric solution
of Cveti\v c, Gibbons, L\"u and Pope (CGLP)~\cite{Cvetic:2000db}. The
perturbed metric and flux coefficients are then functions of only one
radial variable, and generically satisfy $n$ second--order differential
equations.

However, when perturbing around a supersymmetric solution governed by
a superpotential, Borokhov and Gubser~\cite{Borokhov:2002fm} have
observed that these $n$ second--order equations factorize into $n$
first--order equations for certain momenta and $n$ first--order
equations for the metric and flux coefficients, and that furthermore
the $n$ equations for the momenta do not contain the metric and flux
coefficients, and hence can be solved independently. This technique
has been used in several related works~\cite{Bena:2009xk,
  Borokhov:2002fm, Kuperstein:2003yt} and we consider this to be the
technique of choice for deformation problems that depend on just one
coordinate.

\subsection{The first--order Borokhov--Gubser formalism}

While the following summary can be found by now in several sources, we include it here for completeness. 
When the equations of motion governing the fields $\phi^a$ of a certain supersymmetric solution come from the reduction to a one--dimensional Lagrangian
\beq
\mathcal{L} = - \frac{1}{2} G_{ab} \frac{d \phi^a}{ d \tau}  \frac{d \phi^b}{ d \tau}  - V(\phi) \label{Lag1} 
\eeq
whose potential $V(\phi)$ comes from a superpotential,
 \beq
 V(\phi) = \frac{1}{8} G^{ab }\frac{\partial W}{ \partial \phi^a}\frac{\partial W}{ \partial \phi^b}, \label{VWW} \, .
 \eeq
The Lagrangian is written as
 \beq
 \mathcal{L} = - \frac{1}{2} G_{ab} \left( \frac{d \phi^a}{ d \tau}  -\frac{1}{2} G^{ac} \frac{\partial W}{ \partial \phi^c} \right)  \left( \frac{d \phi^a}{ d \tau}  -\frac{1}{2} G^{ac} \frac{\partial W}{ \partial \phi^c}  \right) - \frac{d W}{d \tau} \, ,
 \eeq
and the supersymmetric solutions satisfy
\beq
\frac{d\phi^a}{d\tau}-\frac{1}{2} G^{ab} \frac{\partial W}{\partial \phi^b} =0\label{gradflow} \, .
\eeq

We now want to find a perturbation in the fields $\phi^a$ around their supersymmetric background value $\phi^a_0$ 
  \beq
 \label{split}
 \phi^a=\phi^a_0+\phi^a_1(X) + {\cal O}(X^2)\ ,
 \eeq
where $X$ represents the set of perturbation parameters in which $\phi^a_1$ is linear. 
The deviation from the gradient flow equations for the perturbation $\phi_1^a$ is measured
by the conjugate momenta $\xi_a$
\bea
\xi_a &\equiv& G_{ab}(\phi_0) \left(\frac{d\phi_1^b}{d\tau} -M^b_{\ d} (\phi_0) \phi_1^d \right) \ ,\label{xidef} \\
 M^b{}_d&\equiv&\frac{1}{2} \frac{\partial}{\partial \phi^d} \left( G^{bc} \frac{\partial W}{\partial \phi^c} \right) \ .
\eea
The $\xi_a$ are linear in the expansion parameters $X$, hence they are of the same order as the $\phi_1^a$. When all the $\xi_a$ vanish the deformation is supersymmetric.

The main point of this construction is that the second--order equations of motion governing the perturbations reduce to a set of first--order linear equations for $(\xi_a,\phi^a)$:
\bea
\frac{d\xi_a}{d\tau} + \xi_b M^b{}_a(\phi_0) &=& 0 \, ,  \label{xieq} \\
\frac{d\phi_1^a}{d\tau} - M^a{}_b(\phi_0) \phi_1^b &=& G^{ab} \xi_b \label{phieq} \, .
\eea
Note that equation~\eqref{phieq} is just a rephrasing of the definition of the $\xi_a$ in~\eqref{xidef}, while~\eqref{xieq} implies the equations of motion. Since one considers these perturbations in a metric ansatz in which the reparametrization invariance of the radial variable is fixed, in addition to these equations one must enforce the zero--energy condition
\be \label{ze}
\xi_a \frac{d \phi_0^{a}}{d r} = 0 \, .
\ee

\subsection{The perturbation ansatz}

Using the analysis of the CGLP solution in~\cite{Klebanov:2010qs},
one can easily see that the ansatz for the $\text{SO}(5)$--invariant
eleven--dimensional supergravity solution we are looking for is
\bea\label{ansatz metric}
ds^2 & =& e^{-2z(r)} dx_{\mu} dx^{\mu} + e^{z(r)} \left[ e^{2\, \gamma(r)} \, dr^2 + e^{2 \, \alpha(r)} \sigma_{i}^2 + e^{2\, \beta(r)}\tsig_{i}^2 + e^{2\, \gamma(r)} \nu^2 \right] \nonumber\\
& =& e^{-2z(r)} dx_{\mu} dx^{\mu} + d\tau^2 + a(\tau)^2 \sigma_{i}^2 + b(\tau)^2\tsig_{i}^2 + c(\tau)^2 \nu^2 \, ,  \\
G_4 &=& d K(\tau) \wedge dx^0 \wedge dx^1 \wedge dx^2 + m\, F_4 \ ,
\eea
where $F_4 = d A_3$  and
\bea
A_3 &=& f(\tau)\,\tsig_1 \wedge\tsig_2 \wedge\tsig_3 + h(\tau)\, \epsilon^{ijk}\, \sigma_i \wedge \sigma_j \wedge \tsig_k \\
\Rightarrow F_4 &=& \dot f \, d\tau \wedge \tilde{\sigma}_1 \wedge \tilde{\sigma}_2 \wedge \tilde{\sigma}_3 + \dot h \,\epsilon^{ijk} \, d\tau \wedge \sigma_i \wedge \sigma_j \wedge \tilde{\sigma}_k \nonumber\\ 
&& + \frac{1}{2}\, (4 h - f)\, \epsilon^{ijk}\, \nu \wedge \sigma_i \wedge \tilde{\sigma}_j \wedge \tilde{\sigma}_k - 6\, h\,\nu \wedge \sigma_1 \wedge \sigma_2 \wedge \sigma_3 \, . \label{F4 ansatz}
\eea
Our notation for the one--forms on the Stenzel space is by now standard~\cite{Klebanov:2010qs}, in the sense that with the definitions
\be
\sigma_i = L_{1i} \, ,\ \ \ \ 
\tsig_i = L_{2i} \, ,\ \ \ \ 
\nu = L_{12} \, ,
\ee
they satisfy
\bea\label{d sigma nu L}
d\sigma_i &=& \nu \wedge\tsig_i + L_{ij}\wedge \sigma_j\,,\\
d\tilde{\sigma}_i &=& -\nu \wedge \sigma_i + L_{ij}\wedge\tsig_j\,,\\
d\nu &=& -\sigma_i \wedge \tilde{\sigma_i}\,,\\
dL_{ij} &=& L_{ik} \wedge L_{kj} - \sigma_i \wedge \sigma_j -\tsig_i \wedge\tsig_j\,.
\eea

Integrating one particular component of the equation of motion for the flux
\be
d*G_4 = \half G_4\w G_4
\ee
gives 
\be
\label{H tilde min}
K^{\prime} = 6\, m^2\, \Bslb h\, (f - 2\, h)  - \frac{1}{54} \Bsrb\, e^{-3(\alpha + \beta) - 6 z} \, ,
\ee
where we have chosen the integration constant such that the BPS solution~\cite{Cvetic:2000db}
is regular, i.e.~there are no explicit source M2 branes. We refer to Appendix A for more details.

Performing a standard dimensional reduction on this ansatz down to one dimension, we obtain the following 
Lagrangian
\be
L=(T_{gr}+T_{mat}) - (V_{gr}+V_{mat})
\ee
with the gravitational and matter sectors given by
\bea
T_{gr} &=& 3\, e^{3\,\left(\alpha + \, \beta \right)}\, \left[ \alpha^{\prime 2} + \beta^{\prime 2} - \frac{3}{4}\, z^{\prime 2} + 3 \alpha^{\prime} \beta^{\prime} + \alpha^{\prime} \gamma^{\prime} + \beta^{\prime} \gamma^{\prime} \right] \, , \label{T gr def} \\
V_{gr} &=& \frac{3}{4}\, e^{\alpha + \beta}\, \Bslb e^{4 \alpha} + e^{4 \beta} + e^{4 \gamma} - 2\, e^{2 \alpha + 2 \beta} - 6\, e^{2 \alpha + 2 \gam }\Bsrb \label{V gr def}
\eea
and
\bea
T_{mat} &=& - \frac{m^2}{4}\, e^{3\, \alpha + \beta - 3\, z} \left( f^{\prime 2}\, e^{-4\beta} + 12\, h^{\prime 2}\, e^{-4\alpha} \right)\, , \label{Tmat} \\
V_{mat} &=&\, 3\, m^2\, e^{\alpha + 3\, \beta - 3\, z}\, \left[3\, h^2\,e^{-4\alpha} + \frac{1}{4}\,(4h-f)^2\, e^{-4\beta} \right] \nonumber\\
&& + 9\, m^4\,e^{- 3\, \left( \alpha + \beta + 2\, z\right)}\, \left[ h\, ( f - 2\, h ) - \frac{1}{54}\right]^2 .\label{Vmat}
\eea
The superpotential is given by
\be
W = - 3 \, e^{2 \alpha + 2 \beta} \, (e^{2\alpha} + e^{2\beta} + e^{2\gamma}) - 6\, m^2\, e^{-3 z}\, \left[ h\, (f - 2\, h) - \frac{1}{54} \right].
\ee
It is worth noting that equation~\eq{VWW} only defines the superpotential up one independent minus sign which can then be absorbed in \eq{xieq} and \eq{phieq} by changing the sign of the radial variable and the $\xi_a$. However, with the wisdom of hindsight, we choose a radial variable such that fields decay at infinity and not minus infinity, thus simultaneously fixing the sign of the superpotential.

\subsection{The supersymmetric background}

Here we summarize the expressions that the fields in our ansatz take when
specialized to the zeroth--order CGLP solution~\cite{Cvetic:2000db}
around which we endeavor to study supersymmetric and
non--supersymmetric perturbations.

We should note that the CGLP solution with transverse Stenzel geometry is
to the warped M--theory solution with transverse Stiefel
space~\cite{Ceresole:1999zg} what the IIB Klebanov--Strassler
solution~\cite{Klebanov:2000hb} and the deformed
conifold~\cite{Candelas:1989js} are to the Klebanov--Tseytlin
solution~\cite{Klebanov:2000nc} and the singular conifold. The Stenzel
space is a higher--dimensional generalization of the deformed
conifold. A useful summary of many details of the supergravity solution
can be found in \cite{Martelli:2009ga} and proposals for the dual
field theory can be found in \cite{Martelli:2009ga,Jafferis:2009th}

The supersymmetric solution around which we will perturb was found in~\cite{Cvetic:2000db}. It can be summarized in our ansatz by
\bea\label{0th solt}
e^{2\, \alpha_0} &=& \frac{1}{3}\, \left( 2 + \cosh(2\, r) \right)^{1/4}\, \cosh(r) \, , \\
e^{2\, \beta_0} &=& \frac{1}{3}\, \left( 2 + \cosh(2\, r) \right)^{1/4}\, \sinh(r) \tanh(r) \, ,\\
e^{2\, \gamma_0} &=& \left( 2 + \cosh(2\, r)\right)^{-3/4}\, \cosh^3(r) \, , \\
f_0 &=& \frac{1}{3^{3/2}} \frac{\left( 1 - 3 \cosh^2 (r) \right)}{\cosh^3(r)} \, ,\\
h_0 &=& - \frac{1}{3^{3/2}\, 2} \frac{1}{\cosh(r)} \, ,\\
e^{3z_0(y)} &=&2^{5/2}\, 3\, m^2 \int^{\infty}_{y} \frac{du}{\left( u^4 - 1\right)^{5/2}} \, ,
\eea
where
\beq\label{r to y}
y^4 \equiv 2 + \cosh(2\, r) \, .
\eeq
With this change of coordinate we can write
\bea
e^{3\, z_0} & =&\sqrt{2}\, m^2 \frac{y\, \left(7 - 5\, y^4 \right)}{\left( y^4 - 1\right)^{3/2}} + 5\, \sqrt{2}\, m^2 F\left(\text{arcsin}\left(\frac{1}{y}\right) \mid - 1\right)\, ,
\eea
where the incomplete elliptic integral of the first kind is
\beq\label{F integral}
F(\phi \mid q) = \int_0^{\phi} \left( 1 - q\, \sin(\theta)^2 \right)^{-1/2} d\theta
\eeq
and we have fixed the integration constant (denoted $c_0$ in~\cite{Cvetic:2000db}) by requiring $e^{3z_0}\ra 0$ as $r \ra \infty$.

\subsection{Explicit equations}
We now write out explicitly the two sets of equations \eq{xieq} and \eq{phieq}. In both cases a particular field redefinition simplifies 
things substantially.

\subsubsection{$\xi_a$ equations}

The $\xi^{a}$ equations \eqref{xieq} simplify in the basis
\beq\label{New xi basis}
\tilde{\xi}_a = \left(\xi_1 + \xi_2 + \xi_3, \xi_1 - \xi_2 + 3\, \xi_3, \xi_1 + \xi_2 - 3\, \xi_3,\xi_4, \xi_5, \xi_6 \right)\, .
\eeq
In the order which we solve them, the equations are
\bea
\tilde{\xi}_4^{\prime} &=& 6\, m^2\, e^{-3 (\alpha_0+\beta_0+z_0)}\, \left((f_0 - 2\, h_0)\, h_0 -\frac{1}{54}\right)\, \tilde{\xi}_4 \, ,\label{xi4} \\
\tilde{\xi}_1^{\prime} &=& 12\, m^2\, e^{-3 (\alpha_0+\beta_0+z_0)}\, \left( ( f_0 - 2\, h_0)h_0 -\frac{1}{54} \right)\, \tilde{\xi}_4 \, ,\label{xi1} \\
\tilde{\xi}_5^{\prime} &=& \frac{1}{2}\, e^{\alpha_0-\beta_0}\, \tilde{\xi}_6 - 2\, m^2\, h_0\, e^{-3 (\alpha_0+\beta_0+z_0)}\,\tilde{\xi}_4 \label{xi5} , \\
\tilde{\xi}_6^{\prime} &=&  6\, e^{- 3(\alpha_0 - \beta_0)}\, \tilde{\xi}_5 - 2\, e^{\alpha_0 - \beta_0}\, \tilde{\xi}_6 - 2\, m^2\, e^{-3 (\alpha_0 + \beta_0 + z_0)}\, (f_0 - 4\, h_0)\, \tilde{\xi}_4 \, ,\label{xi6} \\
\tilde{\xi}_3^{\prime} &=& \frac{2}{9}\, e^{-3 (\alpha_0+\beta_0+z_0)}\, \left[18\, e^{2 (\alpha_0+\beta_0+\gamma_0)+3 z_0}\, \tilde{\xi}_3 + m^2\, \left( 54\, h_0\, (f_0 - 2\, h_0) - 1\right)\,\tilde{\xi}_4 \right] \, \non\\&& \label{xi3} \\ 
\tilde{\xi}_2^{\prime} &= & \frac{1}{2}\, e^{-3 \alpha_0 - \beta_0}\, \Big[ 2\, e^{2 (\alpha_0 + \beta_0)} \tilde{\xi}_2 - 6\, e^{2 (\alpha_0+\gamma_0)} \tilde{\xi}_3 - 72\, h_0\, e^{4 \beta_0}\, \tilde{\xi}_5 \nonumber\\
 &&+ e^{4 \alpha_0}\, \left( - 3\, \tilde{\xi}_1 + 2\, \tilde{\xi}_2 + 3\, \tilde{\xi}_3 + 2\, ( f_0\, - 4\, h_0)\, \tilde{\xi}_6 \right) \Big]\, , \label{xi2}
\eea
where we remind the reader that a prime denotes a derivative with respect to $r$ not $y$~\eqref{r to y}.

\subsubsection{$\phi^{a}$ equations} \label{sec:phieq}
The $\phi_a$ equations benefit from a field redefinition as well,
\bea\label{phi redef}
\phi^{a} &=& \left( \alpha, \beta, \gamma, z, f, h \right) \, , \\
\tilde{\phi}_{a} &=& (\phi_1 - \phi_2, \phi_1 + \phi_2 - 2\, \phi_3, \phi_3, \phi_4, \phi_5, \phi_6)\, 
\eea
and we find
\bea
\tilde{\phi}_1^{\prime} &=& \frac{1}{12}\, e^{-3 (\alpha_0+\beta_0)}\, \left[ - 3\, \tilde{\xi}_1 + 4\, \tilde{\xi}_2 + 3\, \left(\tilde{\xi}_3 - 4\, e^{2 (\alpha_0+\beta_0)}\, \left(e^{2 \alpha_0} + e^{2 \beta_0} \right) \tilde{\phi}_1 \right) \right] \, ,\label{phi1} \\
\tilde{\phi}_2^{\prime} &=& \frac{1}{12}\, e^{-3 (\alpha_0 + \beta_0)}\, \left[ - 3\, \tilde{\xi}_1 + 7\, \tilde{\xi}_3 + 12\, e^{2 (\alpha_0 + \beta_0)}\, \left(3\, \left( e^{2 \beta_0} - e^{2 \alpha_0} \right)\, \tilde{\phi}_1 - 4\, e^{2 \gamma_0}\, \tilde{\phi}_2 \right) \right] \, ,\non \\&& \label{phi2}\\
\tilde{\phi}_{3}^{\prime} &=& \frac{1}{12}\, e^{-3 (\alpha_0 + \beta_0)}\, \left[ \tilde{\xi}_1 - 3\,\left( \tilde{\xi}_3 + 6\, e^{2 (\alpha_0+\beta_0)}\, \left( \left( e^{2 \beta_0} - e^{2 \alpha_0} \right)\, \tilde{\phi}_1 - e^{2 \gamma_0}\, \tilde{\phi}_2 \right) \right) \right] \, , \non \\&& \label{phi3}\\
\tilde{\phi}_5^{\prime} &=& \frac{2}{m^2}\, e^{-3 (\alpha_0 - \beta_0)}\, \left[ e^{3 z_0}\, \tilde{\xi}_5 + 3\, m^2\, (3\, h_0\, \tilde{\phi}_1 - \tilde{\phi}_6 ) \right]\, ,\label{phi5} \\
\tilde{\phi}_6^{\prime} &=& \frac{1}{6\, m^2}\, e^{\alpha_0 - \beta_0}\, \left[ e^{3 z_0}\, \tilde{\xi}_6 - 3\, m^2\, (f_0\, \tilde{\phi}_1 - 4\, h_0\, \tilde{\phi}_1 + \tilde{\phi}_5 - 4\, \tilde{\phi}_6 ) \right] \, ,\label{phi6} \\
\tilde{\phi}_4^{\prime}& = &\, \frac{1}{9}\, e^{-3 (\alpha_0 + \beta_0 + z_0)}\, \Big[ 2\, e^{3 z_0}\, \tilde{\xi}_4 + m^2\, \Big( \left[ 1 - 54\, h_0\, (f_0 - 2\, h_0) \right] \tilde{\phi}_4 + 18\, f_0\, \tilde{\phi}_6\nonumber\\ 
&&   + \tilde{\phi}_2+ 2\, \tilde{\phi}_3 + 18\, h_0\, \left[ \tilde{\phi}_5 - 4\, \tilde{\phi}_6 - 3\, ( f_0 - 2\, h_0 )\, (\tilde{\phi}_2 + 2\, \tilde{\phi}_3 ) \right] \Big) \Big] \, . \label{phi4}
\eea

\section{The force on a probe M2}
\setcounter{equation}{0}


Before solving the above equations, we compute the force on a probe
M2--brane in the perturbed solution space. As was found in the
analogous IIB scenario~\cite{Bena:2009xk}, the force turns out to
benefit from remarkable cancellations and is ultimately quite simple.

The membrane action for a probe M2 brane (which by abusing notation we
refer to as the DBI action) is
\begin{align}\label{V DBI}
V^{DBI} & = \sqrt{-\, g_{00}\, g_{11}\, g_{22}} \, ,\nonumber\\
& = e^{-3 z}
\end{align}
and, in the first--order approximation, its derivative with respect to $r$ is
\begin{align}\label{F DBI}
F^{DBI} = - \frac{dV_0^{DBI}}{dr} + 3\, e^{-3 z_0} \left( \tilde{\phi}_4^{\prime} - 3\, z_0^{\prime}\, \tilde{\phi}_4 \right) \, .
\end{align}
We next consider the derivative of the WZ action with respect to $r$, which gives the force exerted on the M2--brane by the $G^{(4)}$ field :
\begin{align}\label{F WZ}
F^{WZ} & = - \frac{dV^{WZ}}{dr} \, , \nonumber\\ 
& = G^{(4)}_{012r}\, , \nonumber\\
& =  - 6\, m^2\, \left[ h\, (f - 2\, h) - \frac{1}{54} \right]\, e^{-3(\alpha + \beta) - 6 z}\, .
\end{align}
The zeroth--order and first--order WZ forces thus are
\beq
F_0^{WZ} =  - 6\, m^2\, \left[ h_0\, (f_0 - 2\, h_0) - \frac{1}{54} \right]\, e^{-3(\alpha_0 + \beta_0) - 6 z_0}\,
\eeq
and
\begin{align}
F_1^{WZ} =&\, - 6\, m^2\, \Big[ h_0\, (\tilde{\phi}_5 - 2\, \tilde{\phi}_6 ) + \tilde{\phi}_6 \, (f_0 - 2\, h_0) \nonumber\\ & - 3\, (\tilde{\phi}_2 + 2\, \tilde{\phi}_3 + 2\, \tilde{\phi}_4)\, \Blp h_0\, (f_0 - 2\, h_0) - \frac{1}{54} \Brp \Big]\, e^{- 3 (\alpha_0 + \beta_0) - 6 z_0}\, .
\end{align}

Combining these two contributions to the force we see that the zeroth--order contributions cancel as expected. Then using the explicit $\phi^a$ equations from section~\ref{sec:phieq} we find the beautiful result
\bea\label{F probe}
F & =& F_1^{DBI} + F_1^{WZ} \nonumber\\
& =& \frac{2}{3}\, e^{-3\,(\alpha_0 + \beta_0 + z_0)(r)}\, \tilde{\xi}_4(r) \, .\nonumber
\eea
At this point it is worthwhile to preemptively trumpet the result \eqref{xi4 integral 2} from Section \ref{sec:solutions} where the exact solution for the mode $\txi_4$ is found:
\bea
F& = & \frac{2}{3}\, e^{-3\,(\alpha_0 + \beta_0)(r)}\, Z_0\, X_4  \nonumber\\
& =& \frac{18\, Z_0\, X_4}{\left( 2 + \cosh2 r \right)^{3/4}\, \sinh^3r} \, ,
\eea
where $Z_0$ is some numerical factor which we found convenient not to absorb into the $X_4$ integration constant,
\beq\label{Zzero}
Z_0 \equiv e^{-3 z_0(0)} \, .
\eeq

So, the UV expansion of the force felt by a probe M2 brane in the
first--order perturbed solution is always 
\beq\label{force UV} F_r
\sim X_4\, e^{-9r/2} + {\cal O}(e^{-17 r/2}) \, .  
\eeq 
In terms of $\rho$, the ``standard'' radial
coordinate\footnote{Related to $r$ via $\cosh(2\, r) \sim
  \rho^{8/3}$.}, this force comes from a potential proportional to
$\rho^{-6}$, which agrees with a straightforward extension of the brane--antibrane force
analysis of~\cite{Kachru:2003sx} to this system. This will be further
discussed in a forthcoming publication~\cite{wip}.

\section{The space of solutions}  \label{sec:solutions}
\setcounter{equation}{0}

In this section we find the generic solution to the
system~\eqref{xi4}--\eqref{phi4}. This solution space has twelve
integration constants of which ten are physical. We have managed to
solve the $\txi_a$ equations exactly whereas for the $\phi_a$
equations we have resorted to solving them in the IR and UV limits.

\subsection{Analytic solutions for the $\tilde{\xi}$'s}

The first equation~\eqref{xi4} is solved by
\beq\label{xi4 integral}
\tilde{\xi}_4 = X_{4}\, \text{exp}\left( 6\, m^2\, \int_{0}^{r} dr^{\prime}\, e^{-3 (\alpha_0+\beta_0+z_0)}\, \left[(f_0 - 2\, h_0)\, h_0 -\frac{1}{54}\right] \right) \, ,
\eeq
which appears to be a double integral. However, using a standard notation for the warp factor $H_0=e^{3z_0}$, since we have
\be
\frac{d H_0}{dr} = - 2^33\, m^2 \frac{e^{2 \gamma_0}}{\sinh^32 r} \tanh^4 r \, ,
\ee
we actually find
\begin{align}\label{xi4 integral 2}
\tilde{\xi}_4 & = X_{4}\, \text{exp}\left( \int_{0}^{r}\, dr^{\prime}\, \frac{1}{H_0} \frac{dH_0}{dr'} \right) \, , \nonumber\\
& = X_{4}\, e^{ 3 (z_{0}(r)-z_0(0) ) } \, .
\end{align}
It immediately follows that
\beq\label{xi1 general solt}
\tilde{\xi}_1 = X_1 + 2\, X_4\, e^{  3 (z_{0}(r)-z_0(0) )} \, .
\eeq
We find convenient not to include $e^{-3 z_0(0)}$ into the integration constant $X_4$, and will use the notation 
\beq
Z_0 \equiv e^{-3 z_0(0)} \, .
\eeq

We were also able to find exact analytic expressions for $\tilde{\xi}_3$ and $\tilde{\xi}_{5,6}$, in term of $y^4 \equiv 2 + \cosh(2\, r)$ :\begin{align}\label{xi3 exact}
\tilde{\xi}_3 =&\, y^4\, \left(y^4-3\right)^2\, X_3 - \frac{m^2\, Z_0\, X_4}{18\, \sqrt{2}}\, \frac{y\, \left( y^4 - 3\right)}{\left( y^4 - 1 \right)^{3/2}}\, \Bigg[- 96 + 599\, y^4 - 550\, y^8 + 119\, y^{12} \nonumber\\ & - y^{3}\, \sqrt{y^4-1}\, \left(3 - 4\, y^4 + y^8 \right)\, \bigg(163\, F\left(\text{arcsin}\left(\frac{1}{y}\right) \mid -1 \right) \nonumber\\ & + 22\, \left[ \Pi \left( -\sqrt{3} ; -\text{arcsin}\left(\frac{1}{y}\right) \mid -1\right) + \Pi \left(\sqrt{3};-\text{arcsin}\left(\frac{1}{y}\right) \mid -1 \right) \right] \bigg) \Bigg] \, , \nonumber\\
\end{align}
where $F(\phi \mid q)$ is given in \eq{F integral} and  $\Pi(n ; \phi \mid m) $ is an incomplete elliptic integral of the third kind
\be
\Pi(n;\phi| m)= \int_{0}^{\phi} \frac{d \theta}{\left(1 - n\, \sin \left( \theta \right)^2 \right)\, \sqrt{1 - m\, \sin \left( \theta \right)^2}}.
\ee

The expressions for $\tilde{\xi}_{5,6}$ are as follows :
\begin{align}\label{xi5 exact}
& \tilde{\xi}_5 = \frac{1}{4\, \sqrt{2}\, \left(y^4 - 3\right)\, \sqrt{y^4 - 1}}\, \Bigg[ \sqrt{6}\, Z_0\, X_4\, m^2\, y\, \left(13 - 11\, y^4 \right)\, \sqrt{y^4 - 1} \nonumber\\ & + 4\, \left[ \left(y^4 - 1\right)^2\, X_5 +\left(y^4 - 3\right)\, \left(1+y^4\right)\, X_6 \right] \nonumber\\ & + \sqrt{6}\, Z_0\, m^2\, X_4\, \bigg[ \left(19 + 7 y^4 \left(y^4 - 2\right) \right)\, F \left( \text{arcsin}\left( \frac{1}{y} \right) \mid -1 \right) \nonumber\\ & - 2\, \left(y^4 - 3\right)\, \left(1+y^4\right)\, \left( \Pi \left( -\sqrt{3}; -\text{arcsin}\left(\frac{1}{y}\right) \mid -1 \right) + \Pi \left( \sqrt{3};-\text{arcsin}\left(\frac{1}{y}\right) \mid -1 \right) \right) \bigg] \Bigg] \, ,\nonumber\\
\end{align}
\begin{align}\label{xi6 exact}
& \tilde{\xi}_6 = \frac{\sqrt{2}}{\left(y^4 - 3\right)\, \left(y^4 - 1\right)^{3/2}}\, \Bigg[ \left(y^4 - 7\right)\, \left(y^4 - 1\right)^2\, \bigg[ X_5 + \sqrt{\frac{3}{2}}\, Z_0\, m^2\, X_4\, \Big( \frac{7\, y - 5\, y^5}{\left(y^4 - 1\right)^{3/2}} \nonumber\\ & + 5\, F \left( \text{arcsin}\left( \frac{1}{y} \right) \mid -1 \right) \Big) \bigg] + \frac{1}{4}\, \left(y^4 - 3\right)^2\, \Bigg[ -\sqrt{6}\, Z_0\, m^2\, X_4\, y\, \sqrt{y^4 - 1} \nonumber\\ & + 4\, \left(y^4 - 3\right)\, X_6 - \sqrt{6}\, Z_0\, m^2\, X_4\, \left(y^4 - 3\right)\, \Bigg(3\, F\left(\text{arcsin}\left( \frac{1}{y} \right) \mid-1 \right) \nonumber\\ & + 2\, \left( \Pi \left(-\sqrt{3}; -\text{arcsin}\left( \frac{1}{y}\right) \mid -1 \right) + \Pi \left( \sqrt{3}; -\text{arcsin}\left( \frac{1}{y} \right) \mid -1 \right) \right)\Bigg) \Bigg] \Bigg] \, .\nonumber\\
\end{align}
Lastly, $\tilde{\xi}_2$ is given by the zero--energy condition \eq{ze} but its explicit form does not appear to be too enlightening.

In Appendix B we provide the IR and UV series expansions of the above solutions for $\tilde{\xi}^{i}$.

\subsection{Solving the $\phi^{i}$ equations}

\subsubsection{The space of solutions}

We now solve the system of equations for $\phi^i$~\eqref{phi1}--\eqref{phi6} using the Lagrange method of variation of parameters. 

Equation~\eqref{phi1} is solved by
\beq\label{phi1 X4 0}
\tilde{\phi}_1 = \frac{\tilde{\lambda}^1(r)}{\sinh(2\, r)}\, ,
\eeq
with
\beq\label{lambda1 X4 0}
\tilde{\lambda}^1 = \frac{9}{2}\, \int\, \frac{\cosh(r)}{\sinh(r)^2\, \left( 2 + \cosh(2\, r) \right)^{3/4}}\, \left[ - 3\, \tilde{\xi}_1 + 4\, \tilde{\xi}_2 + 3\, \tilde{\xi}_3 \right] + Y_1^{IR} \, .
\eeq
$\tilde{\xi}_2$ and $\tilde{\xi}_3$ are given in Section 4.1 above and $\sinh(2\, r)^{-1}$ is the homogeneous solution to the $\tilde{\phi}_1$ equation.

The same Lagrange method is used for $\tilde{\phi}_2$, which is given by
\beq\label{phi2 X4 0}
\tilde{\phi}_2 = \frac{\tilde{\lambda}^2(r)}{\sinh(r)^4\, \left( 2 + \cosh(2\, r) \right)} \, ,
\eeq
where 
\begin{align}\label{lambda2 X4 0}
\tilde{\lambda}^2 =&\, \frac{9}{4}\, \int \, \sinh(r)\, \left( 2 + \cosh(2\, r)\right)^{1/4}\, \left[ - 3\, \tilde{\xi}_1 + 7\, \tilde{\xi}_3 - \frac{4}{3}\, \frac{\sinh(r)^2}{\cosh(r)}\, \left( 2 + \cosh(2\, r) \right)^{3/4}\, \tilde{\phi}_1 \right] \nonumber\\ & \, \, \, + Y_2^{IR} \, .
\end{align}
From this, we obtain an integral expression for $\tilde{\phi}_3$ :
\beq\label{phi3 X4 0}
\tilde{\phi}_3 = \frac{9}{4}\, \int \, \frac{\left[ \tilde{\xi}_1 - 3\, \tilde{\xi}_3 + \frac{2}{3}\, \frac{\sinh(r)^2}{\cosh(r)}\, \left( 2 + \cosh(2\, r) \right)^{3/4}\, \tilde{\phi}_1 + 2\, \frac{\sinh(r)^2\, \cosh(r)^3}{\left( 2 + \cosh(2\, r) \right)^{1/4}}\, \tilde{\phi}_2 \right]}{\sinh(r)^3\, \left( 2 + \cosh(2\, r) \right)^{3/4}} + Y_3^{IR} \, . \nonumber\\
\eeq
The fluxes $\left( \tilde{\phi}_5, \tilde{\phi}_6 \right) = \left( f, h \right)$ are given by
\begin{equation}\label{phi5,6 X4 0}
\left(
\begin{array}{ccc}
\tilde{\phi}_5 \\
\tilde{\phi}_6 \\
\end{array}
\right) = \left(
\begin{array}{ccc}
\cosh(r)^3\, \tanh(r)^6 \, & \cosh(r)^3\, \left[ 2 - 3\, \tanh(r)^2 \right] \\
\frac{1}{2}\, \left[ \text{sech}(r) - \cosh(r)^3 \right] \, & \frac{1}{2}\, \cosh(r)^3 \\
\end{array}
\right)\, \left(
\begin{array}{cc}
\tilde{\lambda}_5 \\
\tilde{\lambda}_6 \\
\end{array}
\right) \, ,
\end{equation}
where the derivatives of $\tilde{\lambda}_5 $ and $\tilde{\lambda}_6 $ are given by 
\begin{equation}\label{lambda5,6 X4 0}
\left(
\begin{array}{ccc}
\tilde{\lambda}_5^{\prime} \\
\tilde{\lambda}_6^{\prime} \\
\end{array}
\right) = \left(
\begin{array}{ccc}
\frac{1}{4}\, \cosh(r)\, \text{coth}(r)^2 \, & \frac{1}{2}\, \left[ \cosh(r) - 2 \text{coth}(r)\, \text{csch}(r) \right] \\
\frac{1}{8}\, \left[ 3 + \cosh(2\, r) \right]\, \text{sech}(r) \, & \frac{1}{2}\, \sinh(r)\, \tanh(r)^3 \\
\end{array}
\right)\, \left(
\begin{array}{cc}
b_5 \\
b_6 \\
\end{array}
\right) \, ,
\end{equation}
and $b_5, b_6$ are the right--hand side of~\eqref{phi5} and~\eqref{phi6} respectively. The $2 \times 2$ matrix appearing in~\eqref{lambda5,6 X4 0} is the inverse of the matrix of homogeneous solutions written in~\eqref{phi5,6 X4 0}. We will call $Y_5$ and $Y_6$ the constants arising from integrating~\eqref{lambda5,6 X4 0}, even though the two functions $\tilde{\phi}_5$ and $\tilde{\phi}_6$ depend on both of them.

Finally, relying on the same method, the equation for $\tilde{\phi}_4$ is solved to
\beq\label{phi4 X4 0}
\tilde{\phi}_4 = e^{-3 z_0(r)}\, \tilde{\lambda}_4 \, , \qquad \tilde{\lambda}_4 = \int\, e^{3 z_0(r)}\, b_4(r) + Y_4^{IR} \, ,
\eeq
where $b_4(r)$ is the right--hand side of~\eqref{phi4} (setting $\tilde{\phi}_4$ to zero). 

\subsubsection{IR behavior}

We now give the IR expansions of the $\phi^{i}$'s. We only write the divergent and constant terms since terms which are regular in the IR do not provide any constraint on our solution space. $Z_0$ is defined in~\eqref{Zzero}.
The $X_{i}$ integration constants are those appearing in the exact solutions for the $\tilde{\xi}_i$'s~\eqref{xi4 integral 2}--\eqref{xi6 exact} :

\begin{align}
\tilde{\phi}_1 =&\, - \frac{1}{r^2}\, \left[ \frac{27\, X_1 + 30\, X_4 - 16\, \sqrt{3}\, X_5 }{4\, 3^{3/4}}\right] +\frac{1}{2\, r}\, Y_1^{IR} \nonumber\\ & + \left[ \frac{ 189\, X_1 + \left( 498 - 198\, 3^{1/4}\, Z_0\, m^2 \right)\, X_4 + 80\, \sqrt{3}\, X_5 }{12\, 3^{3/4}} \right] + {\cal O}(r) \, , \nonumber\\
\end{align}

\begin{align}
\tilde{\phi}_2 =& \, \frac{Y_2^{IR}}{3\, r^4} + \frac{1}{r^2}\, \left[ \frac{9}{4}\, 3^{1/4}\, X_1 + \frac{3}{2}\, 3^{1/4}\, X_4 - 2\, \sqrt{3}\, 3^{1/4}\, X_5 - \frac{4}{9}\, Y_2^{IR} \right] - \frac{1}{2\, r}\, Y_1^{IR} \nonumber\\ & - \left[ 6\, 3^{1/4}\, X_1 + \frac{23}{2}\, 3^{1/4}\, X_4 - 6\, \sqrt{3}\, Z_0\, m^2\, X_4 - \frac{1}{3^{1/4}}\, X_5 - \frac{41}{135}\, Y_2^{IR} \right] \nonumber\\ & + {\cal O}(r) \, , \nonumber\\
\end{align}

\begin{align}
\tilde{\phi}_3 =&\, -\frac{Y_2^{IR}}{8\, r^4} - \frac{1}{r^2}\, \left[ \frac{9\, 3^{1/4}\, X_1 - 12\, 3^{3/4}\, X_5 - 4\, Y_2^{IR} }{24} \right] \nonumber\\ & + \left[ Y_3^{IR} + \frac{3^{1/4}}{8}\, \left(-18\, 3^{1/4}\, Z_0\, m^2\, X_4 + 21\, X_1 + 48\, X_4 + 4\, \sqrt{3}\, X_5 \right)\, \text{log}(r) \right] \nonumber\\ & + {\cal O}(r) \, , \nonumber\\ 
\end{align}

\begin{align}
\tilde{\phi}_4 =&\, - \frac{1}{r^2}\, \left[ \frac{ 18\, X_4 - 4\, \sqrt{3}\, X_5 + Z_0\, m^2\, \left( Y_2^{IR} - 24\, \sqrt{3}\, Y_6^{IR} \right) }{8\, 3^{3/4}} \right] \nonumber\\ & - \bigg[ \frac{1}{4}\, \left( Z_0\, m^2\, \left(\frac{3\, \sqrt{3}}{2}\, X_4 - X_5 \right) - 4\, Z_0\, Y_4^{IR} \right) \nonumber\\ & + \frac{1}{48}\, Z_0^2\, m^4\, \left( \sqrt{3}\, Y_2^{IR} - 72\, Y_6^{IR} \right) + \bigg[ \frac{3}{2}\, 3^{1/4}\, X_4 - \frac{X_5}{3^{1/4}} \nonumber\\ & + \frac{1}{36}\, Z_0\, m^2\, \Big(81\, \sqrt{3}\, X_1 + 78\, \sqrt{3}\, X_4 - 168\, X_5 + 11\, 3^{1/4}\, Y_2^{IR} - 72\, 3^{3/4}\, Y_6^{IR} \Big) \bigg]\, \text{log}(r) \bigg] \nonumber\\ & + {\cal O}(r) \, ,\nonumber\\ 
\end{align}

\begin{align}
\tphi_5 = 2 Y_6^{IR} + \left[ {9 \over 8}\, 3^{3/4}\, X_1 + {3 \over 4}\, 3^{3/4}\, X_4 - 2\, 3^{1/4}\, X_5 + {1 \over 2\, Z_0\, m^2}\, \left( X_5 + {\sqrt{3} \over 2}\, X_4 \right) \right]\, r^2 +  {\cal O}(r^3) ,
\end{align}

\begin{align}
\tilde{\phi}_6 = &\, \frac{1}{r^2}\, \frac{X_5 + \frac{\sqrt{3}}{2}\, X_4}{6\, Z_0\, m^2} \nonumber\\ & + \left[ \frac{3^{3/4}}{16}\, X_1 - \frac{1}{18}\, \frac{X_5 + \frac{\sqrt{3}}{2}\, X_4}{Z_0\, m^2} - \frac{7}{72}\, 3^{3/4}\, X_4 - \frac{5}{18}\, 3^{1/4}\, X_5 + \frac{1}{2}\, Y_6^{IR} \right] + {\cal O}(r) \, .
\end{align}

Note that in the $\tphi_5$ expansion we have also displayed the term of order $r^2$ -- this term will be relevant for the singularity analysis in Section 6.

\subsubsection{UV behavior}

We provide the UV asymptotics for all six $\tilde{\phi}_i$'s, incorporating terms which decay not faster than $e^{-13 r/2}$. However, as appears in Table 1 below, a few modes have leading behavior in the UV which is even more convergent than this.

\begin{align}
\tilde{\phi}_1 = &\, \frac{18}{2^{1/4}}\, X_3\, e^{-r/2} + 2\, Y_1^{UV}\, e^{-2 r} - 4\, 2^{3/4}\, \left[ \frac{27}{2}\, X_1 - 27\, X_3 + 8\, \sqrt{3}\, \left( X_5 + X_6 \right) \right]\, e^{-5 r/2} \nonumber\\ & - \left[ \frac{1089}{10\, 2^{1/4}}\, X_3 - \frac{128}{5}\, 2^{3/4}\, \sqrt{3}\, \left( X_5 + X_6 \right) \right]\, e^{-9 r / 2} + 2\, Y_1^{UV}\, e^{-6 r} \nonumber\\ & + {\cal O}(e^{-13 r / 2}) \, ,
\end{align}

\begin{align}
\tilde{\phi}_2 =&\, \frac{21}{5\, 2^{1/4}}\, X_3\, e^{3 r / 2} - \frac{17523}{140\, 2^{1/4}}\, e^{-5 r/2} X_3 - 12\, Y_1^{UV}\, e^{-4 r} \nonumber\\ & + 4\, 2^{3/4}\, \left[ 99\, X_1 - \frac{1719}{10}\, X_3 +  64\, \sqrt{3}\, \left( X_5 + X_6 \right) \right] \, e^{-9 r / 2} + 32\, Y_2^{UV} e^{-6 r} \nonumber\\ & + {\cal O}(e^{-13 r / 2}) \, ,
\end{align}

\begin{align}
\tilde{\phi}_{3} =&\, - \frac{27}{10\, 2^{1/4}}\, X_3\, e^{3 r / 2} + Y_3^{UV} + \frac{9693}{280\, 2^{1/4}}\, X_3\, e^{-5 r/2} + \frac{15}{4}\, Y_1^{UV}\, e^{-4 r} \nonumber\\ & - 2^{3/4}\, \left[ 130\, X_1 - \frac{1113}{5}\, X_3 + \frac{256}{\sqrt{3}}\, \left( X_5 + X_6 \right) \right]\, e^{-9 r/2} - 12\, Y_2^{UV}\, e^{-6 r} \nonumber\\ & + {\cal O}(e^{-13 r/2}) \, ,\nonumber\\
\end{align}

\begin{align}
\tilde{\phi}_{4} =&\, \frac{3}{16\, 2^{3/4}}\, \frac{Y_4^{UV}}{m^2}\, e^{9 r/2} + \frac{27}{26\, 2^{3/4}}\,\frac{Y_4^{UV}}{m^2}\, e^{5 r/2} + \frac{9}{5\, 2^{1/4}}\, X_3\, e^{3 r/2} + \frac{350271}{183872\, 2^{3/4}}\, \frac{Y_4^{UV}}{m^2}\, e^{r/2} \nonumber\\ & - 2\, \left[ Y_3^{UV} + \sqrt{3}\, \left( Y_5^{UV} - Y_6^{UV} \right) \right] + \frac{216}{325}\, 2^{3/4}\, X_3\, e^{-r/2} + \frac{484605}{298792\, 2^{3/4}}\,\frac{Y_4^{UV}}{m^2}\, e^{-3r/2} \nonumber\\ & + \frac{144}{13}\, \sqrt{3}\, Y_6^{UV}\, e^{-2 r} +\frac{3985953003}{14077700\, 2^{1/4}}\, X_3\, e^{-5 r/2} + \frac{7978373883}{21130570240\, 2^{3/4}}\, \frac{Y_4^{UV}}{m^2}\, e^{-7 r/2} \nonumber\\ & + \left[ \frac{273}{34}\, Y_1^{UV} + \frac{78912\, \sqrt{3}}{2873}\, Y_6^{UV} \right]\, e^{-4 r} \nonumber\\ & - 2^{3/4}\, \left[ 4\, \frac{229}{5}\, X_1 - \frac{1707341851}{2691325}\, X_3 + 4\, \frac{256}{3\, \sqrt{3}} \left( X_5 + X_6 \right) \right]\, e^{-9 r/2} \nonumber\\ & +\frac{473729599251}{995778122560\, 2^{3/4}}\, \frac{Y_4^{UV}}{m^2}\, e^{-11r / 2} + {\cal O}(e^{-6 r}) \, ,\nonumber\\
\end{align}

\begin{align}
\tilde{\phi}_{5} =&\, \frac{1}{8}\, \left(Y_5^{UV} - Y_6^{UV} \right)\, e^{3 r} - \frac{9}{8}\, \left( Y_5^{UV} - Y_6^{UV} \right)\, e^{r} + \frac{1}{8}\, \left( 39\, Y_5^{UV} + 9\, Y_6^{UV} \right)\, e^{-r} \nonumber\\ & + 19\, \frac{4\, 2^{3/4}}{\sqrt{3}}\, X_3\, e^{-3r/2} + \left[ \frac{14}{3\, \sqrt{3}}\, Y_1^{UV} - \frac{1}{8}\, \left(111\, Y_5^{UV} + Y_6^{UV} \right) \right]\, e^{-3 r} \nonumber\\ & - 4\, 2^{3/4}\, \left[ 2\, \frac{279}{65}\, \sqrt{3}\, X_1 + \frac{147}{65}\, \sqrt{3}\, X_3 + 2\, \frac{308}{39}\, \left( X_5 + X_6 \right) \right]\, e^{-7 r/2} \nonumber\\ & + 10\, \left[ - \frac{2}{\sqrt{3}}\, Y_1^{UV} + 3\, Y_5^{UV} \right] e^{-5 r} \nonumber\\ & + \frac{56}{1105}\, 2^{3/4}\, \left[ 3071\, \sqrt{3}\, X_1 - \frac{166409\, \sqrt{3}}{56}\, X_3 + \frac{18716}{3}\, \left( X_5 + X_6 \right) \right]\, e^{-11 r/2} \nonumber\\ & + {\cal O}(e^{-13r/2})  \, ,\nonumber\\
\end{align}

\begin{align}
\tilde{\phi}_{6} =&\, - \frac{1}{16}\, \left( Y_5^{UV} - Y_6^{UV} \right)\, e^{3 r} - \frac{3}{16}\, \left( Y_5^{UV} - Y_6^{UV} \right)\, e^{r} + \frac{1}{16}\, \left( 13\, Y_5^{UV} + 3\, Y_6^{UV} \right)\, e^{-r} \nonumber\\ & + \frac{10}{\sqrt{3}}\, 2^{3/4} X_3\, e^{-3r/2} + \left[ \frac{1}{3\, \sqrt{3}}\, Y_1^{UV} - \frac{1}{16}\, \left( 17\, Y_5^{UV} - Y_6^{UV} \right) \right]\, e^{-3 r} \nonumber\\ & - 4\, 2^{3/4}\,\left[ \frac{33}{65}\, \sqrt{3}\, X_1 + \frac{9\, \sqrt{3}}{130}\, X_3 + \frac{116}{117}\, \left( X_5 + X_6 \right)\right]\, e^{-7 r/2} \nonumber\\ & - \left[ \frac{2}{3\, \sqrt{3}}\, Y_1^{UV} - Y_5^{UV} \right]\, e^{-5 r} \nonumber\\ & + \frac{4}{1105\, \sqrt{3}}\, 2^{3/4}\, \left[ 3713\, X_1 - \frac{30221}{8}\, X_3 + 2932\, \sqrt{3}\, \left( X_5 + X_6 \right) \right]\, e^{-11 r/2} \nonumber\\ & + {\cal O}(e^{-13 r/2}) \, . \nonumber\\
\end{align}

To understand the holographic physics of the $\tilde{\phi}^{i}$ modes, we tabulate the leading UV behavior coming from each mode. To each local operator ${\cal O}_i$ of quantum dimension $\Delta$ in the field theory, the holographic dictionary associates two modes in the dual $AdS$ space, one normalizable and one non--normalizable~\cite{Banks:1998dd, Balasubramanian:1998de}. These two supergravity modes are dual respectively to the vacuum expectation value (VEV) $\langle 0 \mid {\cal O}_i \mid 0\rangle$ and the deformation of the action $\delta S\sim \int d^{d}x {\cal O}_i$:
\begin{align}
\text{normalizable modes}\, \sim \rho_{AdS}^{-\Delta}\, \leftrightarrow \, \text{field theory VEV's} \, \, \, \, \, \, \, \, \, \, \, \, \, \, \, \, \, \, \, \, \, \, \, \, \, \, \, \, \, \, \, \nonumber\\
\text{non--normalizable modes}\, \sim \rho_{AdS}^{\Delta-3}\, \leftrightarrow \, \text{field theory deformations of the action} \, . \nonumber
\end{align}
Here we refer to the standard $AdS$ radial coordinate $\rho_{AdS}$, to be distinguished from the radial coordinate on the cone, $\rho$. In the UV, we have $\rho \sim e^{3 r/4}$ and $\rho_{AdS} \sim \rho^2 / m^{1/3}$ with the factor of $m^{1/3}$ taken with respect to the conventions of~\cite{Klebanov:2010qs}.

In Table 1 we have summarized which integration constants correspond
to normalizable and non--normalizable modes. As stated in a previous
section, the $X_i$ are integration constants for the $\xi_i$ modes and
break supersymmetry, while the $Y_i$ are integration constants for the
modes $\phi^i$. It is very interesting to note that in all cases a
normalizable/non--normalizable pair consists of one BPS mode and one
non--BPS mode.

As already mentioned, the mode $\tilde{\xi}_4$, whose integration
constant is $X_4$ and which is the only mode accountable for the force
felt by a probe M2--brane in the first--order perturbation to the CGLP
background~\cite{Cvetic:2000db}, is the most convergent mode in the
UV, though this cannot be seen from the expansions we have provided
but is apparent at higher order in the asymptotics that we have computed.

\begin{table}[h]\label{UVmodetable}
\onelinecaptionsfalse
\begin{center}
\begin{tabular}{|c|c|c|c|c|c} \hline
dim $\Delta$ & non--norm/norm & int. constant \\ \hline
6 & $\rho_{AdS}^{3}/\rho_{AdS}^{-6}$ & $Y_{4}^{UV}/X_{4}$ \\\hline
5 & $\rho_{AdS}^{2}/\rho_{AdS}^{-5}$ & $Y_{5}^{UV}-Y_{6}^{UV}/X_{5}-X_{6}$ \\\hline
4 & $\rho_{AdS}/\rho_{AdS}^{-4}$ & $X_{3}/Y_{2}^{UV}$ \\\hline
3 & $\rho_{AdS}^0/\rho_{AdS}^{-3}$ & $Y_{3}/X_{2}$ \\\hline
7/3 & $\rho_{AdS}^{-2/3}/\rho_{AdS}^{-7/3}$ & $Y_{5}^{UV}+Y_{6}^{UV}/X_{5}+X_{6}$ \\\hline
5/3 & $\rho_{AdS}^{-4/3}/\rho_{AdS}^{-5/3}$ & $Y_{1}^{UV}/X_{1}$ \\\hline
\end{tabular}
\end{center}
\captionstyle{center}\caption{The UV behavior of the twelve $SO(5)$--invariant modes in the  \protect\\ deformation space of the CGLP solution. As discussed below, only  ten  \protect\\ of these modes are physical, and the mode of dim. 3 is a gauge artifact.}
\end{table}

Taking into account a rescaling which culls $Y_3$ and the zero energy
condition which eliminates $X_2$, we are left with a total of ten
integration constants or five modes. The absence of a physical mode behaving 
as $\rho_{AdS}^0$ is related to the quantization of the
level of the Chern--Simons matter theory.  This is unlike in
four--dimensional gauge theories, where we expect a dimension--four
operator corresponding to the dilaton. Note also that we see explicitly the
dimension $\Delta = 7/3$ operator discussed in~\cite{Klebanov:2010qs}. We have been somewhat glib in
writing $X_5-X_6$ or $Y_5+Y_6$. The numerical factors in the
combination of those integration constants are actually different, but
can be rescaled to the shorthand notation we use.

\section{Boundary conditions for M2 branes} 
\setcounter{equation}{0}

Within the space of solutions that we have derived in Section 4 we now proceed to find the modes which arise from the backreaction of a set of anti--M2 branes smeared on the finite--sized $S^4$ at the tip of the Stenzel-CGLP solution ($r=0$).  For describing them it is necessary to carefully impose the correct infrared boundary conditions.

The gravity solution for a stack of localized M2--branes in flat space has a warp
factor $H(\rho)=1+Q/\rho^6$ and as $\rho \rightarrow 0$ the full solution is smooth due
to the infinite throat. However when these branes are smeared in
$n$--dimensions, the warp factor scales as $\rho^{-6+n}$ as $\rho \rightarrow 0$
since it is now the solution to a wave equation in dimension
$d = 8 - n$. This is the IR boundary condition that we will impose on the
solution.

We must furthermore bring to bear appropriate boundary conditions on the various fluxes.
This is rather simple for M2 branes in flat space, where
the energy from $G^{(4)}$ is the same as that from the curvature. In
the presence of other types of flux, the IR boundary conditions are more intricate. When the background is on--shell, contributions
to the stress tensor from all types of flux taken together cancel the
energy from the curvature: this is the basic nature of Einstein's
equation but this is too wobbly a criterion to signal the presence of
M2 branes. Instead, the right set of boundary conditions for M2 branes should enforce that the dominant contribution to the stress--energy tensor comes from the $G^{(4)}$ flux.

\subsection{BPS M2 branes}

The M2 brane charge varies with the radial coordinate $r$ of a section of the Stenzel space~\cite{Stenzel:1993}:
\begin{align}
\mathcal{Q}_{M2}(r) & = \frac{1}{(2\, \pi\, \ell_p)^6}\, \int_{\mathcal{M}_{7}} \star G_4 \, , \nonumber\\
& = - \frac{6\, m^2\, \text{Vol}\left( V_{5,2} \right)}{\left( 2\, \pi\, \ell_p \right)^6}\, \left( h_0(r)\, \left( f_0(r) - 2\, h_0(r) \right) - \frac{1}{54} \right) \, ,
\end{align}
with $\ell_p$ the Planck length in eleven dimensions, $\mathcal{M}_7$ a constant $r$ section of the transverse Stenzel space of volume $\text{Vol}\left( V_{5,2} \right) = \frac{27\, \pi^4}{128}$~\cite{Bergman:2001qi}.
The number of units of $G_4$ flux through the $S^4$ is
\begin{align}
q(r) & = \frac{1}{\left( 2\, \pi\, \ell_p\, \right)^3}\, \int_{S^4} G_4 \, , \nonumber\\
& = - \frac{16\, \pi^2\, m}{\left( 2\, \pi\, \ell_p \right)^3}\, h_0(r) \, .
\end{align}
In the smooth solution their IR values ($r \rightarrow 0$) are
\beq\label{Q M2 M5 IR}
\mathcal{Q}_{M2}^{\ IR} = 0 \, , \qquad q^{IR} = \frac{1}{\left( 2\, \pi\, \ell_p \right)^3}\, \frac{8\, \pi^2 m}{3^{3/2}} \, , 
\eeq
reflecting the fact that all M2 charge is dissolved in fluxes. One can obtain a BPS solution in which smeared M2 branes are added at the tip of the Stenzel space~\cite{Stenzel:1993} simply by shifting $\star G_4$ in such a way that $f - 4 h$ does not change\footnote{This combination multiplies a four-form field strength with one leg along $\nu$, one along $\sigma^i$ and two legs along two of the $\tilde{\sigma}^j$ directions which shrink in the IR ($e^{2 \beta_0} \sim r^2$)}. Under shifts of $f \rightarrow f + 2\, N$ and $h \rightarrow h + \frac{N}{2}$, the IR M2 brane charge changes to
\beq
{\cal Q}_{M2} \rightarrow {\cal Q}_{M2} + \Delta {\cal Q}_{M2} \, ,
\eeq
where we define
\beq\label{Delta Q M2}
\Delta {\cal Q}_{M2} = - \frac{6\, m^2\, \text{Vol}\left( V_{5,2} \right)}{\left( 2\, \pi\, \ell_p \right)^6}\, \left( \frac{1}{2}\, N^2 - \frac{2}{3^{3/2}}\, N \right) \, ,
\eeq
whereas the variation in the units of flux through the $S^4$ amounts to $\frac{8\, \pi^2\, m\, N}{\left( 2\, \pi\, \ell_p \right)^3}$.
This introduces in the IR a $- \Delta {\cal Q}_{M2} / r^2$ singularity in the warp factor
\beq\
H_0(r) = 162\, m^2\, \int^{r} \frac{h_0\, \left(f_0 - 2 h_0\right) - \frac{1}{54}}{\sinh(r')^3\, \left( 2 + \cosh(2\, r')\right)^{3/4}}\, dr' \, .
\eeq
This singularity is to be expected as we have smeared BPS M2 branes (whose harmonic function diverges as $1/r^6$ near the sources) on the $S^4$ of the transverse space.
It is interesting to see how this BPS solution arises in the first--order expansion around  the BPS CGLP background~\cite{Cvetic:2000db} in the context of our perturbation apparatus. 
Given that the $\xi^i$ modes are associated to supersymmetry--breaking, all the $X_i$ must be set to zero :
\beq
X_i = 0 \, .
\eeq
Since all the $\tilde{\xi}^i$ are zero, 
\beq\label{Y1UVIR}
Y_1^{IR} = Y_1^{UV} \, .
\eeq
In the IR and the UV, $e^{z_0 + 2 \alpha_0}$, $e^{z_0 + 2 \beta_0}$ and $e^{z_0 + 2 \gamma_0}$ do not blow up but reach constant or vanishing values instead. So we impose 
\beq\label{Y12 = 0 BPS}
Y_1^{IR} = 0 \, , \qquad Y_2^{IR} = 0 \, , \qquad Y_4^{UV} = 0 \, .
\eeq 
As a result of~\eqref{Y12 = 0 BPS} and~\eqref{Y1UVIR}, the mode $\tilde{\phi_1}$ is identically zero. This yields $Y_2^{IR} = Y_2^{UV}$, $Y_3^{IR} = Y_3^{UV}$. 

Since BPS M2 branes do not change the geometry of the Stenzel space but only the warp factor (much like BPS D3 branes also only change the warp factor and not the transverse geometry \cite{Grana:2000jj}) we expect the first--order perturbation to $e^{z + 2 \beta}$ to vanish both in the UV and in the IR, and thus
\beq\label{Y3 Y4 Y6 Y5}
2\, Y_3 + e^{-3 z_0(0)}\, Y_4^{IR} + \frac{3}{2}\, m^4\, e^{-6 z_0(0)}\, Y_6^{IR} = 0 \, , \qquad Y_5^{UV} = Y_6^{UV} \, . 
\eeq
The constant $Y_4^{IR}$ is in turn determined by $Y_4^{UV}$.
Furthermore, the fields $\tilde{\phi}_5$, $\tilde{\phi}_6$ now obey the corresponding homogeneous equations and the solution is found by replacing $\tilde{\lambda}_{5,6}$ by $Y_{5,6}$.

The mode $\tilde{\phi}_4$ corresponds to the first--order perturbation of the warp factor. We allow an $1/r^2$ IR divergence, which means that $Y_6^{IR}$ doesn't necessarily need to vanish. We will see in a moment that this mode is related to the number $\Delta {\cal Q}_{M2}$ of added M2 branes. But first, we note that this does not give rise to a singularity that would be associated with $\tilde{\phi}_5 - 4\, \tilde{\phi}_6$, the perturbation to the term in $F_4$~\eqref{F4 ansatz} with legs on $\nu \wedge \sigma_i \wedge\tsig_j \wedge\tsig_k$. Indeed, the conditions we have imposed render this term harmless and independent of $Y_6^{IR}$: $\tilde{\phi}_5 - 4\, \tilde{\phi}_6 = 2\, Y_6 - 2\, Y_6 + {\cal O}(r) = {\cal O}(r)$.

Given that $Y_4^{IR}$ first shows up in the $\mathcal{O}(r^0)$ part of the IR expansion of $\tilde{\phi}_4$ there is no restriction on it. Moreover, $Y_5$ does not arise in any of the divergent or constant pieces in the $\tilde{\phi}^i$ IR expansions, but requiring no exponentially divergent terms in the UV imposes $Y_5 = Y_6$, in agreement with~\eqref{Y3 Y4 Y6 Y5}.

As a result, the perturbation corresponding to adding $\Delta {\cal Q}_{M2}$ M2 branes at the tip is obtained by just setting $Y_5 = Y_6 \sim - \Delta {\cal Q}_{M2}$. This perturbation causes the warp factor to diverge in the infrared as $- \Delta {\cal Q}_{M2} / r^2$ while all the other $\phi^i$ change by sub--leading terms apart from $\phi^5$ and $\phi^6$ which shift by some $N$ related to $\Delta {\cal Q}_{M2}$ through~\eqref{Delta Q M2}.

The UV expansion of the new warp factor is
\begin{align}
H & = e^{3 z_0}\, \left(1 + 3\, \tilde{\phi}_4 \right) \, , \nonumber\\
& = \frac{16}{3}\, 2^{3/4}\, m^2\, e^{-9r/2} \left( 1 - 6\, Y_3 \right) + {\cal O}(e^{-13 r/2}) \, , \nonumber\\
& = \frac{16}{3}\, 2^{3/4}\, m^2\, e^{-9r/2} \left( 1 + 3\, e^{-3z_0(0)}\, Y_4^{IR} + \frac{9}{2}\, m^4\, e^{-6 z_0(0)}\, Y_6 \right) + {\cal O}(e^{-13 r/2}) \, ,
\end{align}
where in the last line we used~\eqref{Y3 Y4 Y6 Y5}, and one can see that $Y_6$ multiplies a $1/\rho^6$ term, as expected from the exact solution.

\section{Constructing the anti--M2 brane solution}
\setcounter{equation}{0}

In order to construct a first--order backreacted solution sourced by anti--M2 branes at the tip of the CGLP solution, the first necessary condition is that the force a probe M2 brane feels be nonzero, which implies:
\beq
X_4 \neq 0 \, .
\eeq
Furthermore, since the infrared is that of a smooth solution perturbed with smeared anti--M2 branes, we require that no other field except those sourced by these anti--M2 branes have a divergent energy density in the infrared. 

Requiring no $1 \over r^2$ or stronger divergences in $\tilde{\phi}_1$, $\tilde{\phi}_2$, $\tilde{\phi}_3$ and $\tilde{\phi}_6$ immediately implies:
\begin{align}
& X_5 = - \frac{\sqrt{3}}{2}\, X_4 \,, \nonumber\\
& Y_2^{IR} = 0 \, , \label{IR-reg}\\
& X_1 = - 2 \, X_4 \, , \nonumber
\end{align}
Barring any $1 \over r$ divergence in $\tilde{\phi}_{1,2}$ results in
\be
Y_1^{IR} = 0 \, .
\ee
The divergence in $\tilde{\phi}_4$ is now
\be
\tilde{\phi}_4 = 3^{1/4}\, {\sqrt{3}\, Z_0\, m^2\, Y_6^{IR} - X_4  \over r^2} + {\cal O}(r^0)
\ee
and this is the proper divergence for the warp factor of anti--M2 branes spread on the $S^4$ in the infrared. The energy density that one can associate with this physical divergence is 
\be
\rho(E)  \sim {d \tilde{\phi}_4 \over d r} \sim {1 \over r^6}
\ee

Another more subtle divergence in the infrared comes from the M--theory four--form field strength, which is 
\bea
G_4 &=& d K(\tau) \wedge dx^0 \wedge dx^1 \wedge dx^2 + m\, F_4 \ ,
\eea
where~\eqref{F4 ansatz}
\bea
F_4 &=& \dot f \, d\tau \wedge \tilde{\sigma}_1 \wedge \tilde{\sigma}_2 \wedge \tilde{\sigma}_3 + \dot h \,\epsilon^{ijk} \, d\tau \wedge \sigma_i \wedge \sigma_j \wedge \tilde{\sigma}_k \nonumber\\ 
&& + \frac{1}{2}\, (4 h - f)\, \epsilon^{ijk}\, \nu \wedge \sigma_i \wedge \tilde{\sigma}_j \wedge \tilde{\sigma}_k - 6\, h\,\nu \wedge \sigma_1 \wedge \sigma_2 \wedge \sigma_3 \ ~.
\eea
The unperturbed metric in the IR is regular and is given by
\beq\label{IR metric}
ds^2 = Z_0^{2/3} \, ds_4^2 + \frac{1}{3^{3/4}} \, Z_0^{-1/3}\, \left[ dr^2 + \nu^2 + \sigma_i^2 + r^2\, \tilde{\sigma}_i^2 \right] \, ,  
\eeq
with the constant $Z_0$ given in~\eqref{Zzero}. The vanishing metric components $g_{\tsig\tsig}$ lead to a
divergent energy density from the four--form field strength components:
\bea
  F_{\nu \sigma \tilde{\sigma} \tilde{\sigma}}  ~F_{\nu \sigma \tilde{\sigma} \tilde{\sigma}} g^{\nu \nu}\, g^{\sigma \sigma}\, g^{\tilde{\sigma} \tilde{\sigma}}\, g^{\tilde{\sigma} \tilde{\sigma}} &= &  \frac{ 9\sqrt{3} Z_0^{4/3} X_4^2}{r^4} +\cO(r^{-2})\\
  F_{r \tsig \tilde{\sigma} \tilde{\sigma}}  F_{r \tsig \tilde{\sigma} \tilde{\sigma}} g^{r r}\, g^{\tsig \tsig}\, g^{\tilde{\sigma} \tilde{\sigma}}\, g^{\tilde{\sigma} \tilde{\sigma}} &=& \frac{81\sqrt{3}Z_0^{3/4}X_4^2}{r^4}+\cO(r^{-2}).
\eea

Unlike the analogous computations in IIB \cite{Bena:2009xk}, when integrating these energy densities the factor of $\sqrt{-G}\sim r^{-3}$ is not strong enough to render the action finite. Hence, this singularity has both a divergent energy density, and a divergent action. 

As discussed in the Introduction, if this singularity is physical then the perturbative solution we find corresponds to the first--order backreaction of a set of anti--M2 branes in the Stenzel-CGLP background. If this singularity is not physical, then our analysis indicates that anti--M2 branes cannot be treated as a perturbation of this background, and hints towards the fact that antibranes in backgrounds with positive brane charge dissolved in fluxes do not give rise to metastable vacua. 

\subsection*{Acknowledgments}
We would like to thank Mariana Gra\~na and Chris Herzog for interesting discussions.
This work is supported in part by a Contrat de Formation par la Recherche of CEA/Saclay, the DSM CEA/Saclay, the grants ANR--07--CEXC--006 and  ANR--08--JCJC--0001--0, and by the ERC Starting Independent Researcher Grant 240210 -- String--QCD--BH. 
\newpage

\begin{appendix}

\section{Subtleties in Section 2.}

To justify our choice of integration constant in~\eqref{H tilde min}, we derive the expression for the non--dynamical scalar $K^{\prime}_0$ in two different ways. First of all, we use the expression~\eqref{H tilde min} for $K^{\prime}$ that arises from its algebraic equation of motion. Inserting the zeroth--order expressions~\eqref{0th solt} of the fields appearing in this expression, we find
\beq\label{H tilde ooth}
K_0^{\prime} = - 3\, m^2\, \frac{\sinh(r)}{\cosh(r)^4}\, \frac{e^{-6 z_0(r)}}{\left( 2 + \cosh(2\, r)\right)^{3/4}}\, .
\eeq

On the other hand, let us proceed to see if this agrees with the expression obtained from the condition that the zeroth--order CGLP solution $K^{\prime}$ has to satisfy
\be
K_0^{\prime} = e^{- 6 z_0(r)}\, \frac{dH_0}{dr} \, ,
\ee 
with $H_0$ solving 
\beq\label{H0 eom}
\nabla_8^2\, H_0 = - \frac{1}{2}\, m^2\, \mid F_4 \mid^2 \, .
\eeq
This reduces to
\begin{align}\label{H0 eom 2}
\frac{dH_0}{dr} = 3\, 2^3\, m^2 \frac{e^{2 \gamma_0}}{\sinh(2\, r)^3} \left( \ell - \tanh(r)^4 \right)\,
\end{align}
and one must set $\ell = 0$ in order for the solution to be regular.
As a result,
\beq\label{K0 2nd}
K_0^{\prime} = - 3\, m^2\, \frac{\sinh(r)}{\cosh(r)^4}\, \frac{e^{-6 z_0(r)}}{\left( 2 + \cosh(2\, r)\right)^{3/4}}\, ,
\eeq
in agreement with the expression for $K_0^{\prime}$ found above from the equation of motion for this non--dynamical field determined in term of $f_0$ and $h_0$~\eqref{0th solt}.

\section{Behavior of $\tilde{\xi}$}

We collect here the infrared and ultraviolet asymptotic expansions of the exact solutions for $\tilde{\xi}^{i}$ which we have derived in Section 4.1.

\subsection{IR behavior of $\tilde{\xi}$}

The IR behavior of the $\tilde{\xi}_a$'s is the following :
\beq
\tilde{\xi}_1^{IR}  = X_1 + 2\, X_4 \left[ 1 - \frac{3^{1/4}}{2}\, m^2\, e^{-3 z_0(0)}\, r^2 \right] + {\cal O}(r^4) \, ,\nonumber
\eeq
\begin{align}
& \tilde{\xi}_2^{IR} = \left[ \frac{3}{2}\, X_1 - \frac{4}{3\, \sqrt{3}}\, X_5 + \frac{7}{3}\, X_4 \right] + \Big[ \frac{3}{2}\, X_1 + \frac{8}{3\, \sqrt{3}}\, X_5 \nonumber\\ & + \frac{1}{3}\, X_4\, \left( 13 - 10\, 3^{1/4}\, e^{-3 z_0(0)}\, m^2 \right) \Big] r^2 + {\cal O}(r^4)\, , \nonumber
\end{align}
\beq
\tilde{\xi}_3^{IR} = 3^{1/4}\, e^{-3 z_0(0)}\, m^2\, X_4\, r^2 + {\cal O}(r^4) \, ,
\eeq
\beq
\tilde{\xi}_4^{IR}  = X_4 \left[ 1 - \frac{3^{1/4}}{2}\, m^2\, e^{-3 z_0(0)}\, r^2 \right] + {\cal O}(r^4) \ , \nonumber
\eeq
\begin{align}
\tilde{\xi}_5^{IR} = &\, \frac{1}{r^2}\, \left[ X_5 + X_4\, \left( \frac{\sqrt{3}}{2} - \frac{3^{3/4}}{2}\, e^{-3 z_0(0)}\, m^2 \right) \right] \nonumber\\ & + \Bigg[ \frac{1}{6}\, \left( 7\, X_5 + 12\, X_6 \right) + X_4\, \bigg[ \frac{17}{20\, \sqrt{3}} - \frac{97}{12}\, 3^{3/4}\, e^{-3 z_0(0)}\, m^2 \nonumber\\ & - \sqrt{6}\, e^{-3 z_0(0)}\, m^2\, \Pi \left(-\sqrt{3};-\text{arcsin}\left(\frac{1}{3^{1/4}}\right) \mid -1 \right)\bigg] - 3^{3/4}\, e^{-3 z_0(0)}\, m^2\, X_4\, \text{log}(r) \Bigg] \nonumber\\ & + \Bigg[ \frac{53}{120}\, X_5 + \frac{1}{48}\, X_4\, \left(\frac{53}{5}\, \sqrt{3} + \frac{47}{5}\, 3^{3/4}\, e^{-3 z_0(0)}\, m^2 \right) \Bigg]  r^2 + {\cal O}(r^4) \, , \nonumber
\end{align}
\begin{align}
\tilde{\xi}_6^{IR} = &\, - \frac{2}{r^2}\, \left[ 2\, X_5 + \sqrt{3}\, X_4 \right] +  \left[ \frac{4}{3}\, X_5 + X_4\, \left( \frac{2}{\sqrt{3}} + 3^{3/4}\, e^{-3 z_0(0)}\, m^2 \right) \right] \nonumber\\ & + \left[ \frac{37}{30}\, X_5 + X_4\, \left( \frac{37}{20\, \sqrt{3}} - 2\, 3^{3/4}\, e^{-3 z_0(0)}\, m^2 \right) \right] r^2 + {\cal O}(r^4) \, . \nonumber
\end{align}

\newpage

\subsection{UV behavior of $\tilde{\xi}$}

The UV behavior of the $\tilde{\xi}_a$'s is as follows :
\beq
\tilde{\xi}_1^{UV} = X_1 + \frac{32}{3}\, 2^{3/4}\, m^2\, X_4\, e^{-3 z_0(0)}\, e^{-\frac{9}{2} r} + {\cal O}(e^{-13 r/2})\, , \nonumber
\eeq
\begin{align}
\tilde{\xi}_2^{UV} = &\, -\frac{3}{32}\, X_3\, e^{6 r} + \frac{3}{16}\, X_3\, e^{4 r} + \left[ \frac{3}{8}\, X_1 + \frac{3}{32}\, X_3 +\frac{2}{3\, \sqrt{3}} \left( X_5 + X_6 \right) \right]\, e^{2 r} \nonumber\\ & + \left[ \frac{3}{4}\, X_1 - \frac{3}{8}\, X_3 -\frac{8}{3\, \sqrt{3}}\,\left( X_5 + X_6 \right) \right] \nonumber\\ & + \left[ \frac{3}{8}\, X_1 + \frac{3}{32}\, X_3 + \frac{2}{3\, \sqrt{3}}\, \left( X_5 + X_6 \right) \right]\, e^{-2 r} \nonumber\\ & + \left[ \frac{3}{16}\, X_3 + \frac{64}{3\, \sqrt{3}}\, X_6 \right]\, e^{-4 r} + \frac{32}{7}\, 2^{3/4}\, e^{-3 z_0(0)}\, m^2\, X_4\, e^{-9r/2} \nonumber\\ & - \left[ \frac{3}{32}\, X_3 + \frac{256}{3\, \sqrt{3}}\, X_6 \right]\, e^{-6 r} + {\cal O}(e^{-13 r/2}) \, , \nonumber
\end{align}
\begin{align}
\tilde{\xi}_3^{UV} = &\, \frac{1}{8}\, X_3\, e^{6 r} - \frac{9}{8}\, X_3\, e^{2 r} + 2\, X_3 - \frac{9}{8}\, X_3\, e^{-2 r} \nonumber\\ & + \frac{32}{7}\, 2^{3/4}\, e^{-3 z_0(0)}\, m^2\, X_4\, e^{-9r/2} + \frac{1}{8}\, X_3\, e^{-6 r} + {\cal O}(e^{-13 r/2})\, , \nonumber 
\end{align}
\beq
\tilde{\xi}_4^{UV} = \frac{16}{3}\, 2^{3/4}\, m^2\, X_4\, e^{-3 z_0(0)}\, e^{-\frac{9}{2} r} + {\cal O}(e^{-13 r/2})\, ,
\eeq
\begin{align}
\tilde{\xi}_5^{UV} =&\, \frac{1}{2}\, \left( X_5 + X_6 \right)\, e^{r} + \frac{5}{2}\, \left( X_5 + X_6 \right)\, e^{-r} + 2\, \left( 3\, X_5 - X_6 \right)\, e^{-3 r} \nonumber\\ & + 2\, \left( 5\, X_5 + X_6 \right)\, e^{-5 r} - \frac{96}{13}\, 2^{3/4}\, \sqrt{3}\, e^{-3 z_0(0)}\, m^2\, X_4\, e^{-11r/2} + {\cal O}(e^{-13r/2}) \, , \nonumber
\end{align}
\begin{align}
\tilde{\xi}_6^{UV} =&\, \left( X_5 + X_6 \right)\, e^{r} - 7\, \left( X_5 + X_6 \right)\, e^{-r} - 24\, \left( X_5 - X_6 \right)\, e^{-3 r} \nonumber\\ & - 8\, \left( 5\, X_5 + 7\, X_6 \right)\, e^{-5 r} - \frac{192}{13}\, 2^{3/4}\, \sqrt{3}\, m^2\, X_4\, e^{-3 z_0(0)}\, e^{-11r/2} + {\cal O}(e^{-13 r/2}) \, . \nonumber
\end{align}

\end{appendix}

\providecommand{\href}[2]{#2}\begingroup\raggedright\endgroup
\end{document}